\documentclass[12pt,english]{article}
\pdfoutput=1
\usepackage[T1]{fontenc}
\usepackage{graphicx,array}
\usepackage{cite}
\usepackage{color}
\usepackage{latexsym}
\usepackage{amsthm}
\usepackage{amsmath}
\usepackage[titletoc]{appendix}
\usepackage{enumitem}
\usepackage{amssymb}
\usepackage{color}
\usepackage[unicode=true,
 bookmarks=true,bookmarksnumbered=false,bookmarksopen=false,
 breaklinks=false,pdfborder={0 0 1},backref=false,colorlinks=true]
 {hyperref}
\hypersetup{pdftitle={Area Law Unification and the Holographic Event Horizon},
 pdfauthor={Yasunori Nomura, Grant N. Remmen},
 citecolor=black,linkcolor=black,urlcolor=black}
\usepackage{breakurl}
\usepackage[hang,flushmargin]{footmisc} 

\def\simgt{\mathrel{\lower2.5pt\vbox{\lineskip=0pt\baselineskip=0pt
           \hbox{$>$}\hbox{$\sim$}}}}
\def\simlt{\mathrel{\lower2.5pt\vbox{\lineskip=0pt\baselineskip=0pt
           \hbox{$<$}\hbox{$\sim$}}}}

\newcommand{\be}{\begin{equation}}
\newcommand{\ee}{\end{equation}}
\newcommand{\bea}{\begin{eqnarray}}
\newcommand{\eea}{\end{eqnarray}}
\newcommand{\Eq}[1]{Eq.~(\ref{#1})}
\newcommand{\Eqs}[2]{Eqs.~(\ref{#1}) and (\ref{#2})}
\newcommand{\Sec}[1]{Sec.~\ref{#1}}
\newcommand{\Secs}[2]{Secs.~\ref{#1} and \ref{#2}}
\newcommand{\Fig}[1]{Fig.~\ref{#1}}

\newcommand{\Ref}[1]{Ref.~\cite{#1}}
\newcommand{\Refs}[2]{Refs.~\cite{#1,#2}}

\newcommand{\So}[1]{S^{({\rm outer})}[#1]}

\newcommand{\scri}{{\cal I}}

\newcommand*\oline[1]{%
  \vbox{%
    \hrule height 0.5pt%                  % Line above with certain width
    \kern0.68ex%                          % Distance between line and content
    \hbox{%
      \kern-0.1em%                        % Distance between content and left side of box, negative values for lines shorter than content
      \ifmmode#1\else\ensuremath{#1}\fi%  % The content, typeset in dependence of mode
      \kern-0.1em%                        % Distance between content and left side of box, negative values for lines shorter than content
    }% end of hbox
  }% end of vbox
}

\definecolor{nicered}{rgb}{0.7,0.1,0.1}
\definecolor{nicegreen}{rgb}{0.1,0.5,0.1}
\hypersetup{
 colorlinks,citecolor=black,linkcolor=black,urlcolor=black}
\usepackage{xcolor}

\begin{document}

\interfootnotelinepenalty=10000
\baselineskip=18pt
\hfill

\vspace{2cm}
\thispagestyle{empty}
\begin{center}
{\LARGE \bf
Area Law Unification\\[2mm] and the Holographic Event Horizon
}\\
\bigskip\vspace{1cm}{
{\large Yasunori Nomura and Grant N. Remmen}
} \\[7mm]
{\it Center for Theoretical Physics and Department of Physics \\
     University of California, Berkeley, CA 94720, USA and \\
     Lawrence Berkeley National Laboratory, Berkeley, CA 94720, USA}
\let\thefootnote\relax\footnote{e-mail: 
\url{ynomura@berkeley.edu}, \url{grant.remmen@berkeley.edu}}
 \end{center}

\bigskip
\centerline{\large\bf Abstract}
\begin{quote} \small
We prove a new, large family of area laws in general relativity, which 
apply to certain classes of untrapped surfaces that we dub generalized 
holographic screens.  Our family of area laws contains, as special 
cases, the area laws for marginally-trapped surfaces (holographic 
screens) and the event horizon (Hawking's area theorem).  In addition 
to these results in general relativity, we show that in the context 
of holography the geometry of a generalized holographic screen is 
related to the outer entropy of the screen.  Specifically, we show 
for spherically-symmetric spacetimes that the area of the largest 
HRT surface consistent with the outer wedge can be computed in terms 
of the geometry of the general (not necessarily marginally-trapped) 
codimension-two surface defining the wedge.  This outer entropy 
satisfies a second law of thermodynamics, growing monotonically 
along the generalized holographic screen.  In particular, this 
result provides the holographic dual for the geometry of the event 
horizon for spherically-symmetric spacetimes.
\end{quote}
	
\setcounter{footnote}{0}

\newpage
\tableofcontents
\newpage

\section{Introduction}
\label{sec:Introduction}

Area laws for dynamical surfaces in spacetime have, both historically 
and recently, been important drivers of progress in theoretical physics. 
Under certain positivity conditions for the flow of energy-momentum, 
general relativity constrains the dynamics of certain surfaces such that 
their area only increases.  The most well known example is Hawking's 
area law for black holes~\cite{Hawking:1971tu,Bardeen:1973gs}, which 
mandates that the area of the event horizon always grows with time. 
This provided the basis for the thermodynamic understanding of black 
holes~\cite{Bardeen:1973gs,Bekenstein:1972tm,Bekenstein:1973ur,%
Hawking:1974rv,Hawking:1974sw}; in turn, black hole thermodynamics 
helped inspire the development of holography~\cite{tHooft:1993dmi,%
Susskind:1994vu,Bousso:2002ju}.  In the context of string theory, 
the AdS/CFT correspondence~\cite{Maldacena:1997re,Gubser:1998bc,%
Witten:1998qj,Aharony:1999ti} has provided the prime example of a 
tractable holographic model that can be explored in detail.  Holography 
thus gives us powerful tools with which to understand quantum gravity.

The areas of extremal surfaces in asymptotically-AdS spacetimes have 
proved to be of significance beyond their geometrical interpretation. 
They correspond to entanglement entropies of regions in the boundary 
CFT, given by the Ryu-Takayanagi formula~\cite{Ryu:2006bv,Ryu:2006ef,%
Lewkowycz:2013nqa} for static slices and more generally by the 
Hubeny-Rangamani-Takayanagi (HRT) prescription~\cite{Hubeny:2007xt,%
Wall:2012uf,Dong:2016hjy}.  An understanding of the dynamics of these 
surfaces can shed light on the entanglement structure of the boundary 
and vice versa, with the question being actively researched from both 
the gravitational and field theory perspectives~\cite{Wald:1993nt,%
VanRaamsdonk:2010pw,Maldacena:2013xja,Dong:2013qoa,Faulkner:2013ica,%
Bao:2015nqa,Bao:2015nca,Remmen:2016wax,Bao:2017thr}.

Building on earlier work~\cite{Hayward:1993wb,Hayward:2004fz}, 
an interesting area theorem in general relativity was recently
proved~\cite{Bousso:2015mqa,Bousso:2015qqa,Sanches:2016pga} for 
holographic screens, a substantive extension of apparent horizons 
to timelike or spacelike objects whose slices are marginally-trapped 
or -antitrapped surfaces~\cite{Bousso:1999cb}.  Such screens can be 
found in many spacetimes of interest, such as expanding universes 
and inside of black holes, and it was shown that these surfaces have 
areas that grow in a particular direction along the screen.  While 
geometrically interesting in their own right, such surfaces---as their 
name implies---have been suggested to have a holographic interpretation, 
as the surfaces on which to formulate a ``boundary'' theory in 
general spacetimes beyond AdS~\cite{Nomura:2016ikr,Nomura:2017npr,%
Nomura:2017fyh,Nomura:2018kji}, though at present no explicit boundary 
theory is known for this more general conjectured form of holography. 
Moreover, an entropic interpretation of the area of the holographic 
screen has been demonstrated~\cite{Engelhardt:2017aux}: the area 
of an apparent horizon equals the area of the largest HRT surface 
compatible with the domain of dependence of the spacetime outside 
the apparent horizon.  That is, the apparent horizon area can be 
viewed as an ``outer entropy'' of the spacetime.  In contrast, 
despite the success of Hawking's area theorem in sparking black 
hole thermodynamics and the holographic revolution, a valid 
holographic interpretation of the event horizon itself has remained 
elusive~\cite{Hubeny:2012wa,Engelhardt:2017wgc}.

In this paper, we will show that both the holographic screen and 
the event horizon are special cases of a much more general class of 
surfaces, which we will call {\it generalized holographic screens}, 
all of which satisfy an area law.  Thus, we will unify the area law 
discovered in \Refs{Bousso:2015mqa}{Bousso:2015qqa} and Hawking's area 
law~\cite{Hawking:1971tu}.  These generalized holographic screens extend 
the concept of holographic screens to surfaces that are not marginally 
trapped; these new surfaces sweep out large portions of the interior 
of a black hole and can also be constructed in cosmological spacetimes. 
These results are proved purely in general relativity and are 
independent of holography.

Furthermore, we will show, for spherically-symmetric spacetimes, that 
the outermost spacelike portion of generalized holographic screens 
have an entropic interpretation analogous to that of apparent horizons 
given in \Ref{Engelhardt:2017aux}.  In particular, we will prove another 
new general relativity result, giving the area of the largest HRT surface 
compatible with the outer wedge of a slice of the generalized holographic 
screen.  The area of this maximal HRT surface is given by a geometric 
quantity computable in terms of the area and curvature of the generalized 
holographic screen.  Viewed as a holographic statement, we compute the 
outer entropy of a non-marginally-trapped surface inside a black hole. 
This implies that we find a new entry in the holographic dictionary: 
the entropic interpretation of the event horizon (in terms of its area 
and curvature), for spherically-symmetric spacetimes.  Comparing the 
evolution of the maximal HRT area associated with different slices, 
we show that the outer entropy satisfies a second law, despite being 
a complicated function of geometric quantities on the generalized 
holographic screen.

The remainder of this paper is organized as follows.  In \Sec{sec:GHS}, 
we define our terminology and give the definition of generalized 
holographic screens, in particular proving in \Sec{subsec:area} that 
they satisfy an area law.  In \Sec{sec:outer}, we review the definition 
of outer entropy and show that, for the generalized holographic screen, 
it is upper bounded by the area of the screen in Planck units.  In 
\Sec{sec:sphere} we compute the outer entropy for spherically-symmetric 
spacetimes; we discuss several special cases of interest in 
\Sec{subsec:cases} and prove the second law for the outer entropy 
in \Sec{subsec:2nd-law}.  We conclude and discuss future directions 
in \Sec{sec:concl}.

\section{Generalized Holographic Screens}
\label{sec:GHS}

In this section, we derive our results based on classical general 
relativity.  First, we will discuss some differential geometry formalism 
and review the notion of (marginally-trapped) holographic screens. 
We will then introduce the notion of generalized holographic screens 
and establish our family of area laws, illustrating how Hawking's area 
theorem for event horizons arises as a special case.

\subsection{Formalism and Review}
\label{subsec:formalism}

Throughout the paper, we will consider a smooth spacetime 
$(M,g_{ab})$ of dimension $D \geq 3$ that is globally hyperbolic 
(or, in the asymptotically-AdS case, with appropriate boundary 
conditions~\cite{Avis:1977yn}).  We will also assume the Einstein 
equations and the null energy condition (NEC), $T_{ab} k^a k^b \geq 0$ 
for energy-momentum tensor $T_{ab}$ and any null vector $k^a$; 
equivalently, we could assume the null curvature condition (NCC) 
$R_{ab} k^a k^b \geq 0$.  We will use mostly-plus metric signature 
and sign conventions $R_{ab} = R^{c}_{\;\;acb}$ and $R^{a}_{\;\;bcd} 
= \partial_c \Gamma^a_{bd} - \partial_d \Gamma^a_{bc} + \Gamma^a_{ce} 
\Gamma^e_{bd} - \Gamma^a_{de} \Gamma^e_{bc}$.  We follow the standard 
differential geometry notation, defining the chronological future 
(respectively, past) of a set $S$ as $I^\pm (S)$, the future 
(respectively, past) domains of dependence $D^\pm (S)$ as the 
set of points $p \in M$ such that every past (respectively, future) 
inextendible causal curve through $p$ in $M$ intersects $S$, and 
the domain of dependence $D(S)$ as the union $D^+(S) \cup D^-(S)$. 
We use a dot $\dot{S}$, circle $\mathring{S}$, and bar $\overline S$ to 
denote the boundary, interior, and closure of a set $S$, respectively. 
In our conventions, $S \not \subset I^\pm (S)$, but $S \subset 
D^\pm (S)$.

Let us first review some results of \Ref{Bousso:2015qqa}.  We define 
a {\it future holographic screen} $H$ to be a smooth (codimension-one) 
hypersurface for which one can define a {\it foliation} (i.e., a partition 
of $H$) into marginally-trapped codimension-two compact acausal surfaces 
called {\it leaves}.  From a leaf $\sigma$, we will call the two 
future-directed orthogonal null geodesic congruences $k$ and $l$; 
the marginally-trapped condition stipulates that, on $\sigma$,
\be
  \theta_k = 0 \qquad \text{and} \qquad \theta_l < 0,
\ee
where $\theta_k = \nabla_a k^a$ and $\theta_l = \nabla_a l^a$ are the 
null expansions for $k$ and $l$, respectively.  Defining an area element 
$\delta A$, the expansions can equivalently be written as $\theta_k 
= \nabla_k \log \delta A$ and $\theta_l = \nabla_l \log \delta A$, where 
$\nabla_k = k^a \nabla_a$ and $\nabla_l = l^a \nabla_a$ are the covariant 
derivatives along the congruences.  Throughout, we will extend the 
definition of $k$ and $l$ to null vector fields over the entire spacetime 
$M$.  For a given $H$, this choice of $k$ and $l$ over all of $M$ is 
not unique, but our results will hold for all such choices.

We consider the case in which each leaf $\sigma$ splits some Cauchy 
surface $\Sigma$ into two disjoint subsets, $\Sigma = \Sigma^+ \cup 
\sigma \cup \Sigma^-$, where $\sigma = \dot \Sigma^\pm$ and we label 
$\Sigma^-$ as the outer portion (which we take to be in the $k$ 
direction) and $\Sigma^+$ as the inner portion (which we take to 
be in the $l$ direction); we choose this notation and the $\pm$ 
convention for outer versus inner to match that of \Ref{Bousso:2015qqa}. 
We note that $\sigma$ divides the spacetime into four disjoint portions, 
$I^\pm (\sigma)$ and $D(\Sigma^\pm)$; in particular, $I^\pm(\sigma)$ 
and $\overline D(\Sigma^\pm) - \sigma$ together constitute a four-part 
partition of $M-\sigma$, as shown in Fig.~2 of \Ref{Akers:2017nrr}. 
As proved in \Ref{Akers:2017nrr}, the boundaries of these regions can 
be characterized by the geodesic congruences $k$ and $l$, truncating 
at any conjugate points (i.e., caustics) or intersections of 
finitely-separated geodesics.  This fact will be used frequently 
in our arguments that follow.

One can define a real parameter $\tau$ on $H$ such that each leaf 
$\sigma$ is a surface of constant, unique $\tau$.  We can also write 
the tangent vector field $h^a$ parallel to the leaf-orthogonal curves 
within $H$ as
\be
  h^a = \alpha l^a + \beta k^a 
\ee
for some real parameters $\alpha$ and $\beta$, normalized so that 
$h^a (d\tau)_a = 1$.

\begin{figure}[t]
\centering
  \includegraphics[width = 12cm]{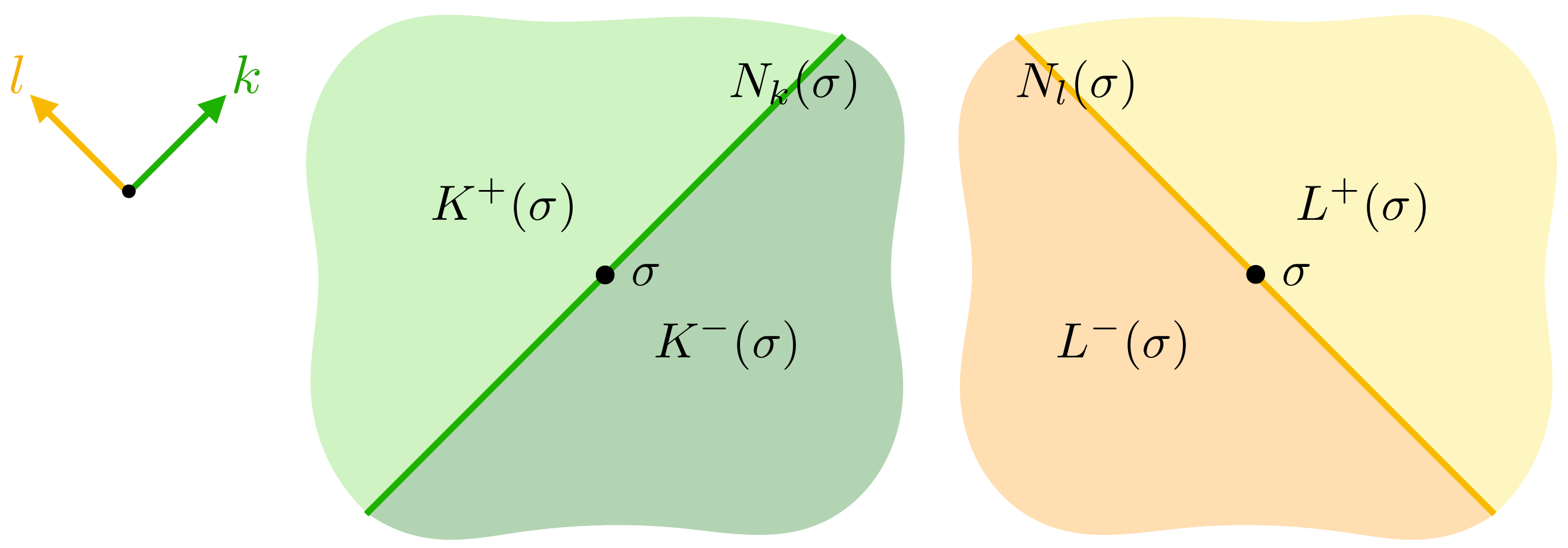}
\caption{Generic Penrose diagrams for spacetime regions $K^\pm(\sigma)$ 
 and $L^\pm(\sigma)$ defined in \Eq{eq:KL}, divided by the light sheets 
 $N_k(\sigma)$ and $N_l(\sigma)$, respectively, defined in \Eq{eq:Nkl}.}
\label{fig:KL}
\end{figure}

We can then make the following definitions of null surfaces:
\be
\begin{aligned}
  N_{+k}(\sigma) &= \dot I^+ (\Sigma^+) - \Sigma^+ 
    = \dot D^+(\Sigma^-) - I^-(D^+(\Sigma^-)) \\
  N_{-k}(\sigma) &= \dot I^-(\Sigma^-) - \Sigma^- 
    = \dot D^-(\Sigma^+) - I^+(D^-(\Sigma^+)) \\
  N_{+l}(\sigma) &= \dot I^+(\Sigma^-) - \Sigma^- 
    = \dot D^+(\Sigma^+) - I^-(D^+(\Sigma^+)) \\
  N_{-l}(\sigma) &= \dot I^-(\Sigma^+) - \Sigma^+ 
    = \dot D^-(\Sigma^-) - I^+(D^-(\Sigma^-)).
\end{aligned}
\label{eq:Nkl}
\ee
The result of \Ref{Akers:2017nrr} implies that the expressions on 
the right-hand side are independent of the choice of Cauchy surface 
$\Sigma$ and are indeed defined only by the leaf $\sigma$; that is, 
$N_{\pm k}$ and $N_{\pm l}$ are {\it light sheets}, null surfaces 
defined up to caustics and nonlocal intersections of null geodesics. 
We further define $N_k(\sigma) = N_{+k}(\sigma) \cup N_{-k}(\sigma)$ 
and $N_l(\sigma) = N_{+l}(\sigma)\cup N_{-l}(\sigma)$ and note that 
$\sigma = N_{+k}(\sigma) \cap N_{-k}(\sigma) = N_{+l}(\sigma) \cap 
N_{-l}(\sigma)$.  Given the Cauchy-surface-independence, we define 
the spacetime regions
\be
\begin{aligned}
  K^+(\sigma) &= I^+(\Sigma^+) \cup D^-(\Sigma^+) - N_{-k}(\sigma) \\
  K^-(\sigma) &= I^-(\Sigma^-) \cup D^+(\Sigma^-) - N_{+k}(\sigma) \\
  L^+(\sigma) &= I^+(\Sigma^-) \cup D^-(\Sigma^-) - N_{-l}(\sigma) \\
  L^-(\sigma) &= I^-(\Sigma^+) \cup D^+(\Sigma^+) - N_{+l}(\sigma),
\end{aligned}
\label{eq:KL}
\ee
so $N_k(\sigma) = \dot K^+(\sigma) = \dot K^-(\sigma)$ and $N_l(\sigma) 
= \dot L^+(\sigma) = \dot L^-(\sigma)$; see \Fig{fig:KL}.

Finally, as in \Ref{Bousso:2015qqa} we will take $R_{ab} k^a k^b + 
\varsigma_k^2$ to be strictly positive on $H$, where $\varsigma_k$ is the 
shear tensor of the $k$ congruence as defined in \Ref{Wald}.  Along with 
the Raychaudhuri equation
\be
  \nabla_k \theta_k = -\frac{1}{D-2}\theta_k^2 
    - \varsigma_k^2 - R_{ab} k^a k^b
\label{eq:Raychaudhuri}
\ee
and the NEC, this genericity assumption implies that $\theta_k$ is 
strictly positive (negative) to the past (respectively, future) of 
$\sigma$.  Note that the term involving the twist tensor is absent 
in \Eq{eq:Raychaudhuri} because the congruence is surface-orthogonal.

Given these conditions and additional technical assumptions,%
\footnote{\Ref{Bousso:2015qqa} also assumes that every inextendible 
 portion of $H$ contains either a complete leaf or is completely 
 timelike and that the sets of points in $H$ for which $\alpha$ is 
 positive and negative share a boundary on which $\alpha$ vanishes.}
\Ref{Bousso:2015qqa} then shows that $\alpha < 0$ everywhere on $H$. 
That is, $h^a$ points either (timelike) to the past or (spacelike) 
outwards.  As a result, the sets of $K^\pm$ are monotonic under 
inclusion: writing $K^\pm(\tau) = K^\pm(\sigma(\tau))$, one obtains 
the inclusion relations
\be
\begin{aligned}
  \overline K^+(\tau_1) &\subset K^+(\tau_2) 
    &\qquad(\alpha<0,\text{any }\beta) \\
  \overline K^-(\tau_2) &\subset K^-(\tau_1) 
    &\qquad(\alpha<0,\text{any }\beta) \\
\end{aligned}
\label{eq:inclusionk}
\ee
for $\tau_1 < \tau_2$.  Analogously, if we can choose a region 
where $\beta$ is constant throughout a leaf, with the same sign 
at $\sigma(\tau_1)$ and $\sigma(\tau_2)$, the sets $L^\pm(\tau) 
= L^\pm(\sigma(\tau))$ are also monotonic under inclusion:
\be
\begin{aligned}
  \overline L^+(\tau_2) &\subset L^+(\tau_1) &&\qquad(\alpha<0,\beta>0) \\
  \overline L^-(\tau_1) &\subset L^-(\tau_2) &&\qquad(\alpha<0,\beta>0) \\
  \overline L^+(\tau_1) &\subset L^+(\tau_2) &&\qquad(\alpha<0,\beta<0) \\
  \overline L^-(\tau_2) &\subset L^-(\tau_1) &&\qquad(\alpha<0,\beta<0).
\end{aligned}
\label{eq:inclusionl}
\ee
Finally, \Ref{Bousso:2015qqa} shows that the holographic screen $H$ 
satisfies an area law: $A[\sigma(\tau_1)] < A[\sigma(\tau_2)]$, so 
$dA/d\tau > 0$.%
\footnote{Throughout, we will use round brackets for scalar arguments. 
 For objects that take a set of points in $M$ as an argument, we will 
 use round brackets if the object being defined is itself a subset 
 of the spacetime (e.g., $D(S)$), while we will use square brackets 
 in the case of a quantity defined on the spacetime (e.g., $A[S]$ for 
 the area of a surface $S$).}
By reversing the time direction and swapping past for future in all of 
the definitions, one can define {\it past holographic screens}, which 
are foliated by marginally-antitrapped surfaces and which also satisfy 
an area law.

\subsection{Definition of Generalized Holographic Screens}
\label{subsec:def}

\begin{figure}[t]
\centering
  \includegraphics[width = 11cm]{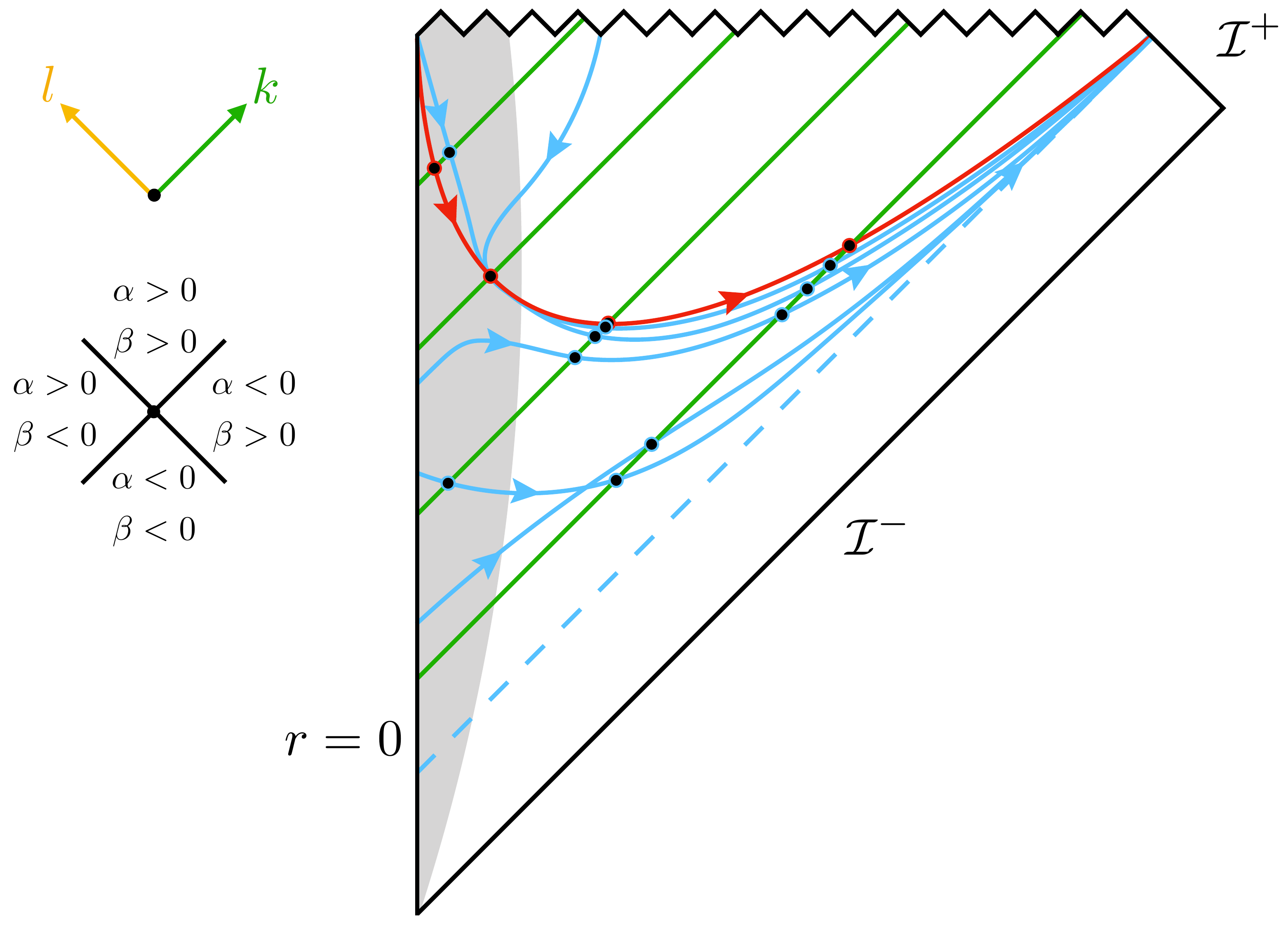}
\caption{Generalized (future) holographic screens inside a black hole 
 formed from collapse.  The holographic screen $H$ is shown in red 
 and has both a timelike and spacelike portion.  Several examples 
 of generalized holographic screens $H'$ are illustrated by the blue 
 curves.  In this example, they can have both timelike and spacelike 
 portions or can be purely spacelike.  The limiting case of the 
 event horizon (blue dashed line) also corresponds to a generalized 
 holographic screen.  A few representative light sheets $N_k(\sigma)$ 
 are illustrated by the green lines and on these light sheets the 
 codimension-two leaves $\sigma$ of $H$ (on which $\theta_k = 0$) 
 and the leaves $\sigma'$ of $H'$ (on which $\theta_k \neq 0$) are 
 represented by the black dots; for each $\sigma' \subset H'$ there 
 exists $\sigma \subset H$ for which $\sigma' \subset N_k(\sigma)$. 
 We will show that all of these screens obey an area law, with 
 increasing area toward the past and outward directions, as illustrated 
 by the arrows pointing in the direction of increasing $\tau$.}
\label{fig:examples}
\end{figure}

We will show that there is a much larger family of surfaces, beyond 
the holographic screens discussed in \Sec{subsec:formalism}, that also 
satisfy an area law.  In particular, we are interested in relaxing 
the requirement that the leaves be marginally trapped.  Given a future 
holographic screen $H$ as described in \Sec{subsec:formalism}, we 
will define a {\it generalized future holographic screen} $H'$ as a 
surface to the past (future) of $H$ when $H$ is spacelike (respectively, 
timelike), with $H'$ being spacelike if and only if the corresponding 
section of $H$ is spacelike.  A few examples of generalized future 
holographic screens are shown in \Fig{fig:examples}.  We will later 
prove that $H'$ satisfies an area law, but before that let us first 
specify the conditions defining $H'$ more precisely.

Formally, we define a generalized future holographic screen as a 
(codimension-one) hypersurface $H'$ with a foliation into codimension-two 
leaves $\sigma'$ and tangent vector $h'^a = \alpha l^a + \beta k^a$ 
(for some $\alpha$ and $\beta$) satisfying the following criteria:
\begin{enumerate}
\item For each $\sigma' \subset H'$, there exists $\sigma \subset H$ 
for which $\sigma' \subset N_k(\sigma)$.  For each $p \in \sigma$, we 
can identify some $p' \in \sigma'$ satisfying $p' \in N_k(p)$.  If $h$ 
is spacelike or null, then we require $p' \in N_{-k}(p)$, while if $h$ 
is timelike or null, $p' \in N_{+k}(p)$.
\label{cond1}
\item The signature and orientation of $h'$ at $p' \in \sigma'$ matches 
that of $h$ at $p \in \sigma$ for which $p' \in N_k(p)$.
\label{cond2}
\item $\theta_l < 0$ on $H'$.
\label{cond3}
\end{enumerate}

For {\it generalized past holographic screens}, $N_{\pm k}$ are 
simply swapped in condition~\ref{cond1}, while condition~\ref{cond3} 
becomes $\theta_l > 0$.  By \Eq{eq:inclusionk}, $N_k(\sigma(\tau_1)) 
\cap N_k(\sigma(\tau_2)) = \varnothing$ for $\tau_1 \neq \tau_2$, so 
the $\sigma$ for which $\sigma' \subset N_k(\sigma)$ is unique.  That 
is, there is a function $\phi: \mathbb{R} \rightarrow \mathbb{R}$ for 
which $\sigma'(\tau) \subset N_k(\sigma(\phi(\tau))$.  Note that $\phi$ 
is not necessarily injective or surjective: there may be more than one 
slice $\sigma' \subset H'$ in the same $N_k(\sigma)$ and there may be 
some $\sigma \subset H$ for which $N_k(\sigma) \cap H' = \varnothing$.

Note that the event horizon itself is a generalized future holographic 
screen, corresponding to the limit in which $\phi(\tau)$ maps all 
numbers to infinity, where the leaves of the original holographic 
screen $\sigma(\tau)$ go to $\scri^+$ as $\tau \rightarrow \infty$. 
While the event horizon is teleologically defined (i.e., it requires 
knowledge of the entire future history of the spacetime), the 
holographic screen is defined quasilocally, in terms of metric 
and its derivatives measurable at a point, in a particular Cauchy 
slicing.  The generalized holographic screen shares characteristics 
of both of these definitions: it is defined in terms of the holographic 
screen, but using past- or future-directed light sheets.  Hence, 
the event horizon is a generalized holographic screen in the particular 
limit in which all $\sigma'$ are in $N_{-k}(\sigma)$ for the leaf 
$\sigma = H \cap \scri^+$ on the boundary of the spacetime.

\subsection{Area Law}
\label{subsec:area}

We now show that there is an area law on the generalized holographic 
screen $H'$.  By condition~\ref{cond1} in \Sec{subsec:def}, for 
the region where $H$ is timelike (respectively, spacelike), we have 
$\theta_k < 0$ (respectively, $\theta_k > 0$) on $H'$ by the Raychaudhuri 
equation~\eqref{eq:Raychaudhuri} and the fact that $\theta_k = 0$ on 
$H$.  By condition~\ref{cond2}, we thus have $\theta_k \leq 0$ when 
$H'$ is timelike (or null) and $\theta_k \geq 0$ when $H'$ is spacelike 
(or null).

Moreover, condition~\ref{cond2} implies that $\alpha < 0$ on $H'$, 
since $\alpha < 0$ on $H$.  That is, $h'^a$ is either past- or 
outward-directed, so $\beta < 0$ when $H'$ is timelike and $\beta > 0$ 
when $H'$ is spacelike.  Hence, $\beta \theta_k \geq 0$ on $H'$.  By 
condition~\ref{cond3}, $\theta_l < 0$ on $H'$, so $\alpha \theta_l > 0$. 
That is, we have shown that the general covariant definition of $H'$ 
given in conditions~\ref{cond1} through~\ref{cond3} implies
\be
  \alpha \theta_l + \beta \theta_k > 0
\label{eq:alphabeta}
\ee
everywhere on $H'$.

\begin{figure}[t]
\centering
  \includegraphics[width = 7.5cm]{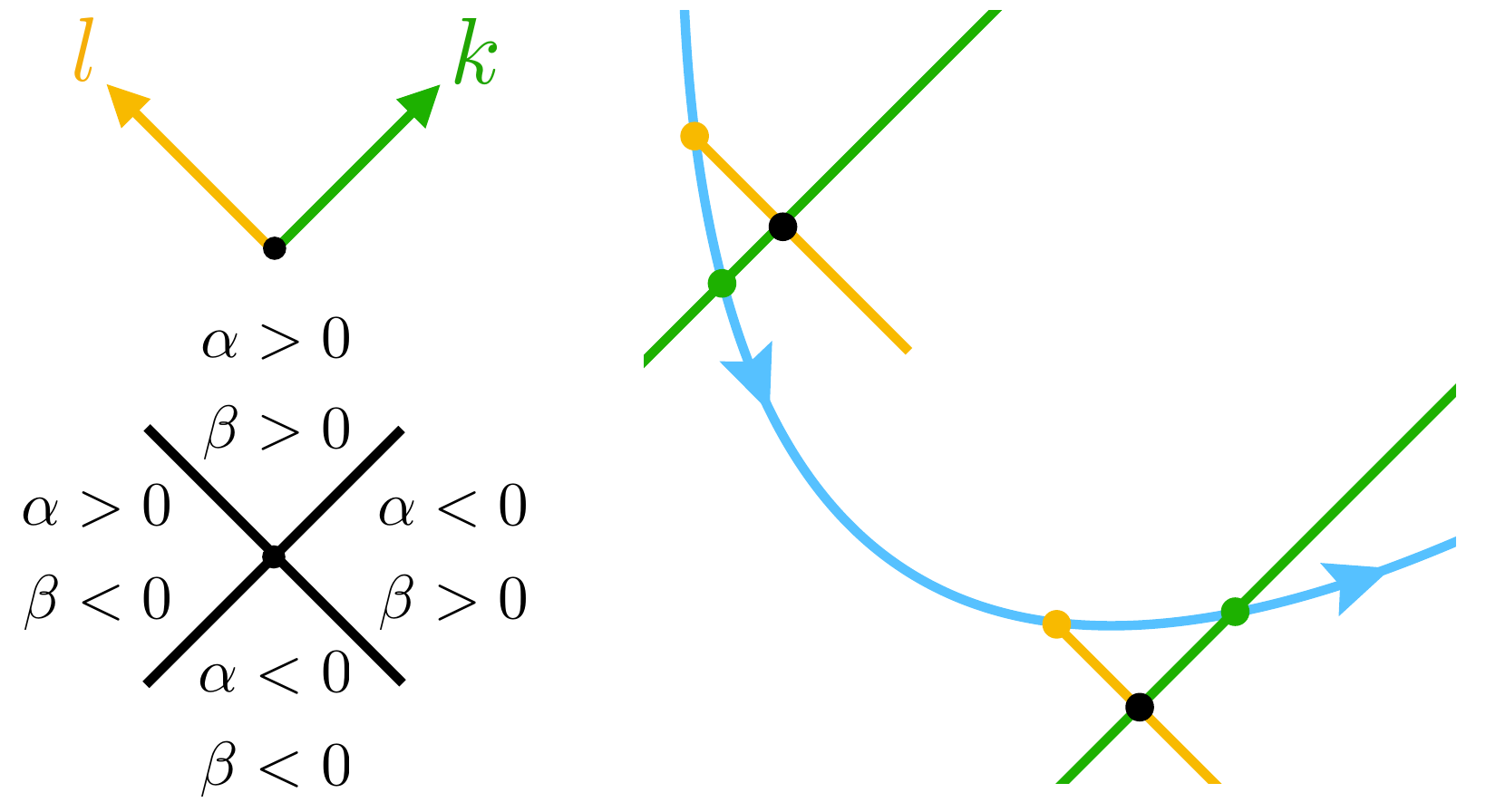}
\caption{Zigzag construction proving the area law on a generalized 
 holographic screen $H'$ (blue curve).  From $\sigma'(\tau)$ 
 (examples given by orange dots), we take the past $l$ light 
 sheet $N_{-l}(\sigma'(\tau))$ (orange lines), while from 
 $\sigma'(\tau+d\tau)$ (green dots), we take the $k$ light 
 sheet $N_k(\sigma'(\tau+d\tau))$ (green lines).  The intersection 
 $\tilde\sigma(\tau,\tau+d\tau)$ (black dots) is always nonempty. 
 We have $A[\tilde\sigma(\tau,\tau+d\tau)] > A[\sigma'(\tau)]$ and 
 $A[\sigma'(\tau+d\tau)] > A[\tilde\sigma(\tau,\tau+d\tau)]$, so 
 area increases along $H'$.}
\label{fig:zigzag}
\end{figure}

We can now adapt the zigzag argument of \Ref{Bousso:2015qqa} to prove 
an area law on $H'$.  Let us first consider the case in which $\sigma'$ 
is smooth; we will subsequently extend our result to the more general 
case of non-smooth $\sigma'$.  Since $\theta_l < 0$ on $H$, by continuity 
there always exists a surface near $H$ with $\theta_l < 0$ satisfying 
conditions \ref{cond1} and \ref{cond2}, so choices of $H'$ always 
exist.  Given smooth $\sigma$, there always exists a smooth surface 
$\sigma' \subset N_k(\sigma)$ by taking $\sigma'$ sufficiently 
near $\sigma$, since by a theorem of \Ref{Hawking:1973uf}, 
geodesics cannot exit the boundary of the future or past of 
$\sigma$ instantaneously.  From $\sigma'(\tau) \subset H'$, 
consider the light sheet $N_{-l}(\sigma'(\tau))$ going in the 
past $l$ direction.  From $\sigma'(\tau+d\tau) \subset H'$, 
consider the null hypersurface $N_k(\sigma'(\tau+d\tau))$, following 
both the past and future $k$ directions.  Since $\alpha < 0$, 
$\tilde{\sigma}(\tau,\tau+d\tau) = N_{-l}(\sigma'(\tau)) \cap 
N_{k}(\sigma'(\tau+d\tau))$ is nonempty; see \Fig{fig:zigzag}. 
For regions of $H'$ that are spacelike ($\beta > 0$), 
$\tilde{\sigma}(\tau,\tau+d\tau) \subset N_{-k}(\sigma'(\tau+d\tau))$. 
Conversely, for parts of $H'$ that are timelike ($\beta < 0$), 
$\tilde{\sigma}(\tau,\tau+d\tau) \subset N_{+k}(\sigma'(\tau+d\tau))$.

Since $\tilde{\sigma}(\tau,\tau+d\tau) \subset N_{-l}(\sigma'(\tau))$, 
we have
\be
  A[\tilde{\sigma}(\tau,\tau+d\tau)] - A[\sigma'(\tau)] 
  = A[\tilde{\sigma}(\tau,\tau+d\tau)] \alpha \theta_l d\tau
\label{eq:dA1}
\ee
for infinitesimal $d\tau$, recalling the definition of $\theta_l 
= \nabla_l \log\delta A$.  Similarly, the change in area from 
$\tilde{\sigma}(\tau,\tau+d\tau)$ to $\sigma'(\tau+d\tau)$ is
\be
  A[\sigma'(\tau+d\tau)] - A[\tilde{\sigma}(\tau,\tau+d\tau)] 
  = A[\tilde{\sigma}(\tau,\tau+d\tau)] \beta \theta_k d\tau
\label{eq:dA2}
\ee
since $\theta_k = \nabla_k \log\delta A$.  Hence,
\be
  A[\sigma'(\tau+d\tau)] - A[\sigma'(\tau)] 
  = A[\tilde{\sigma}(\tau,\tau+d\tau)] 
    (\alpha \theta_l + \beta \theta_k) d\tau.
\ee
By \Eq{eq:alphabeta}, we therefore have
\be
  A[\sigma'(\tau+d\tau)] - A[\sigma'(\tau)] > 0,
\ee
leading to an area law along $H'$:
\be
  \frac{dA[\sigma'(\tau)]}{d\tau} > 0.
\label{eq:arealawfinal}
\ee
Specifically, writing the induced metric on $\sigma'(\tau)$ as 
$\gamma_{ab}^{\sigma'(\tau)}$, the area grows at the rate
\be
  \frac{dA[\sigma'(\tau)]}{d\tau} = \int_{\sigma'(\tau)} 
    \sqrt{\gamma^{\sigma'(\tau)}} 
    (\alpha \theta_l[\sigma'(\tau)]+ \beta \theta_k[\sigma'(\tau)]).
\label{eq:dAcalc}
\ee
We thus have a general covariant geometric formulation of a generalized 
holographic screen that is not a marginally-trapped surface but that 
nonetheless satisfies an area law.

Let us now generalize our result by relaxing the requirement that 
$\sigma'$ is so close to $\sigma$ as to be smooth.  In particular, 
$\sigma'$ can now contain portions of caustics or nonlocal 
intersections in $N_k(\sigma)$, where null geodesics can 
enter or exit the light sheet defining the past or future of 
$\sigma$~\cite{Akers:2017nrr}.  Even in this case, an area 
law can be proved.  For a spacelike part of $H'$, between 
$\tilde\sigma(\tau,\tau+d\tau)$ and $\sigma'(\tau+d\tau)$, 
it is possible for future-directed null geodesics to enter 
$N_{-k}(\sigma'(\tau+d\tau))$, but not to leave it (see, e.g., 
Refs.~\cite{Hawking:1973uf,Wald}).  Similarly, for a timelike 
part of $H'$, it is possible for future-directed null geodesics 
to leave but not enter $N_{+k}(\sigma'(\tau+d\tau))$ between 
$\sigma'(\tau+d\tau)$ and $\tilde\sigma(\tau,\tau+d\tau)$.  Hence, 
in both cases $A[\sigma'(\tau+d\tau)]-A[\tilde\sigma(\tau,\tau+d\tau)]$ 
is lower-bounded by the right-hand side of \Eq{eq:dA2} and is 
therefore still positive.

We next consider the other light sheets defining 
$\tilde\sigma(\tau,\tau+d\tau)$, i.e.\ $N_{-l}(\sigma'(\tau))$. 
Future-directed null geodesics cannot leave $N_{-l}(\sigma'(\tau))$; 
however, they can enter $N_{-l}(\sigma'(\tau))$ only when they 
encounter a caustic or a nonlocal intersection with a null geodesic 
originating from elsewhere on $\sigma'$ \cite{Akers:2017nrr}. 
If they entered through a caustic, one would find that, moving from 
past to future, their expansion $\theta_l$ jumps discontinuously 
from $-\infty$ to $+\infty$ at the entry point and then decreases 
continuously toward $\sigma'(\tau)$.  This implies that since 
$\theta_l$ is by definition negative on $\sigma'(\tau)$, we can 
always choose $d\tau$ sufficiently small that a caustic is not 
encountered between $\sigma'(\tau)$ and $\tilde\sigma(\tau,\tau+d\tau)$ 
on $N_{-l}(\sigma'(\tau))$.  It is also clear that $d\tau$ can always 
be chosen small enough that the generators of $N_{-l}(\sigma'(\tau))$ 
do not encounter any nonlocal intersections between $\sigma'(\tau)$ 
and $\tilde\sigma(\tau,\tau+d\tau)$.  Hence, \Eq{eq:dA1} still holds, 
and the right-hand side of \Eq{eq:dAcalc} gives a lower bound on 
the rate of area increase.  We thus find that the area increase rate 
is still positive.  Namely, we have an area law on the generalized 
holographic screen $H'$ even if $\sigma'$ is not close to $\sigma$.

The original holographic screen $H$ is a special case of our family of 
generalized holographic screens $H'$, taking the limit in which $\sigma' 
\rightarrow \sigma$ for all $\tau$, so that $\theta_k \rightarrow 0$. 
Hence, the area law for $H'$ reduces smoothly to the area law for the 
holographic screen $H$ derived in \Ref{Bousso:2015qqa}.

Moreover, Hawking's area theorem~\cite{Hawking:1971tu} is also a special 
case of our area law for generalized holographic screens.  In the case 
of a holographic screen $H$, the marginally-trapped condition prescribes 
a particular foliation into leaves $\sigma$.  For a region of a 
generalized holographic screen $H'$ where the mapping between leaves 
$\sigma' \subset H'$ and $\sigma \subset H$ is one-to-one (i.e., the 
function $\phi$ is injective), $H'$ inherits the foliation of $H$. 
However, if we choose $H'$ to have a finite null region, then multiple 
leaves in $H'$ lie within $N_k(\sigma)$ for the same $\sigma \subset H$. 
In this region, the foliation of $H$ does not prescribe a foliation 
of $H'$; under any foliation of a null portion of $H'$ into leaves 
$\sigma'$, the area law proved above still applies by virtue of 
the positivity of $\theta_k$.  Similarly, Hawking's area theorem is 
independent of the spacelike Cauchy slicing: for any two spacelike 
Cauchy slices $\Sigma_1$ and $\Sigma_2$ where $\Sigma_2 \subset 
I^+(\Sigma_1)$, the event horizon $\dot I^-(\scri^+)$ grows in area, 
so $A[\dot I^-(\scri^+) \cap \Sigma_1]\leq A[\dot I^-(\scri^+)\cap 
\Sigma_2]$~\cite{Hawking:1973uf}.  Hence, for any spacelike Cauchy 
slicing of the spacetime, we can define a foliation of a null portion 
of $H'$ simply via its intersection with the Cauchy slices.  In 
Hawking's area theorem, the area law follows from proving that the 
expansion on the horizon is nonnegative in a spacetime satisfying the 
NCC.  In our present context, assuming an asymptotically-stationary 
spacetime, so that the horizon is asymptotically marginally trapped, 
implies that there exists a holographic screen $H$ that asymptotes to 
the horizon.  We can thus define the horizon itself as a generalized 
holographic screen $H'$, on which $\theta_k$ is positive by the 
Raychaudhuri equation~\eqref{eq:Raychaudhuri}.

Our family of generalized holographic screens thus unifies two 
previously known area laws associated with black holes, namely, 
those of the holographic screen and the event horizon.  This 
unification is nontrivial: while it is true that a convex combination 
of two monotonic functions is itself monotonic, such intuition does 
not readily apply to spacetime geometries, in which the notion of 
taking a combination of two surfaces is not in general well defined 
without specifying additional geometric information for how to 
determine the new surface.  Our definition in \Sec{subsec:def} 
provides precisely the requisite specifications, guaranteeing, 
as we have shown in this section, an area law for the generalized 
holographic screen.

\subsection{Alternate Construction of Screens}
\label{subsec:intersection}

The definition of generalized holographic screens in \Sec{subsec:def} 
leads to immense freedom in choosing $H'$.  The only requirements 
are those given in conditions~\ref{cond1} through~\ref{cond3}.

\begin{figure}[t]
\centering
  \includegraphics[width = 10cm]{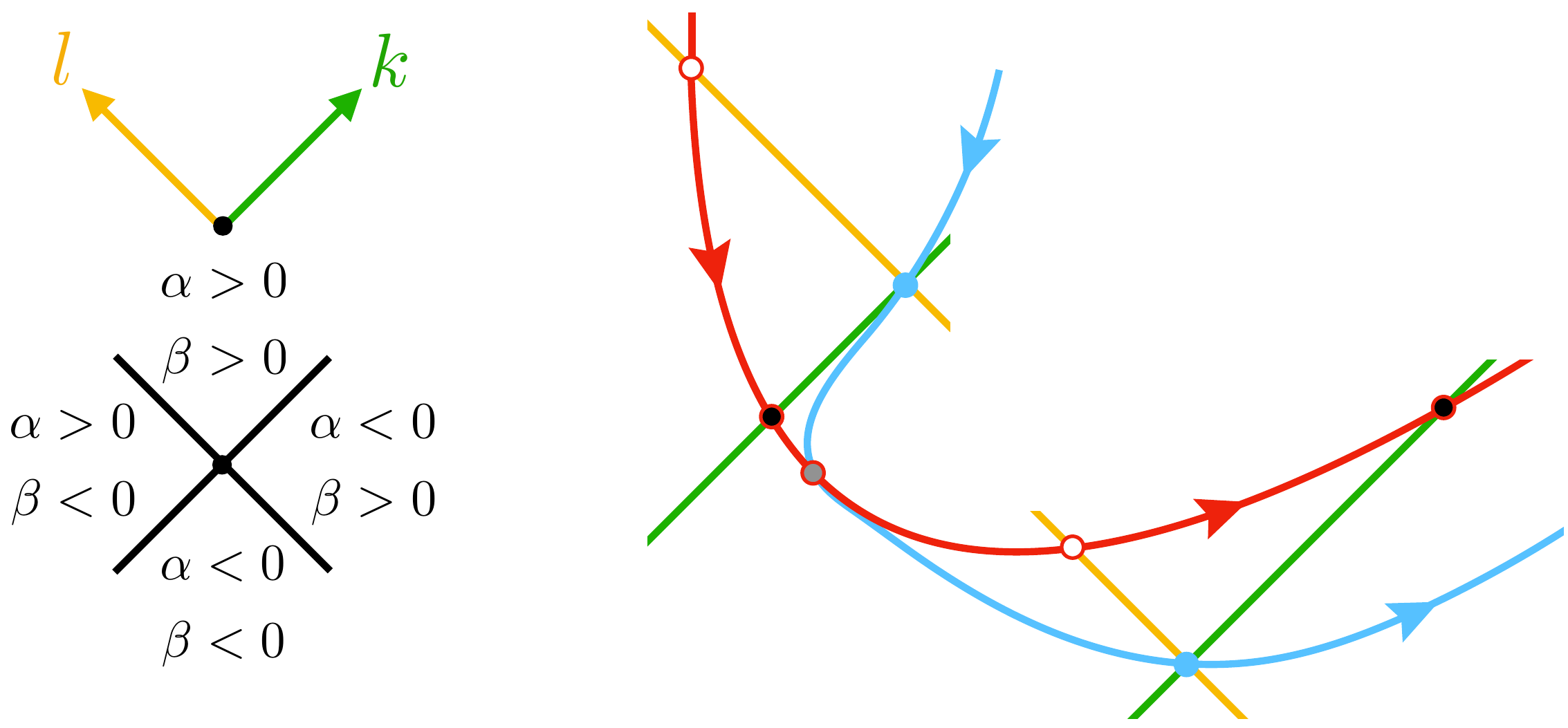}
\caption{Illustration of the intersection construction of generalized 
 holographic screens.  Arrows indicate the direction of increasing 
 $\tau$.  Examples of leaves $\sigma(\tau)$ in the holographic screen 
 $H$ (red curve) are given by the white dots and the corresponding 
 $\sigma(f(\tau))$ is given by the immediately succeeding black dot. 
 A leaf $\sigma'$ (blue dot) of the generalized holographic screen 
 $H'$ (blue curve) is given by the intersection of $N_l(\sigma(\tau))$ 
 (yellow line) and $N_k(\sigma(f(\tau)))$ (green line) as shown in 
 \Eq{eq:leavesalt}.  The function $f(\tau) \geq \tau$ equals $\tau$ 
 precisely when $H$ has null tangent (gray dot).}
\label{fig:GHS_alt}
\end{figure}

However, we can formulate an elegant alternative way of defining 
a particular subset of generalized future holographic screens 
parameterized by a single real function.  Let $f:\, \mathbb{R} 
\rightarrow \mathbb{R}$ be a smooth function with $df/d\tau > 0$ 
and $f(\tau) \geq \tau$, with equality if and only if $\sigma(\tau) 
\subset H$ has null tangent $h^a$.  In this subsection, we will 
also assume for simplicity that each leaf of $H$ is entirely timelike, 
spacelike, or null.  Then we can remove conditions~\ref{cond1} 
and~\ref{cond2} and instead simply define $H'$ to be the hypersurface 
foliated by leaves
\be
  \sigma'(\tau) =  N_l(\sigma(\tau)) \cap N_k(\sigma(f(\tau))).
\label{eq:leavesalt}
\ee
We still require condition~\ref{cond3} that $\theta_l < 0$.  See 
\Fig{fig:GHS_alt} for an illustration of this construction.  The 
analogous construction for generalized past holographic screens can 
be defined similarly.

On the spacelike part of $H$ (on which $\beta > 0$), for $\tau_1 
< \tau_2$, \Eq{eq:inclusionl} implies $\sigma'(\tau_2) \subset 
N_{-l}(\sigma(\tau_2)) \subset \overline L^+(\sigma(\tau_2)) 
\subset L^+(\sigma(\tau_1))$, while by \Eq{eq:inclusionk}, 
$\sigma'(\tau_2) \subset N_{-k}(\sigma(f(\tau_2))) \subset 
\overline K^-(\sigma(f(\tau_2))) \\ \subset K^-(\sigma(f(\tau_1)))$, 
since $f(\tau_1) < f(\tau_2)$ by definition of $f$.  Now, for any 
cross section $\hat\sigma$ of $N_k(\sigma)$, $K^{\pm}(\sigma) = 
K^{\pm}(\hat\sigma)$, while for any cross section $\hat\sigma$ 
of $N_l(\sigma)$, $L^{\pm}(\sigma) = L^{\pm}(\hat\sigma)$. 
Hence, $K^-(\sigma(f(\tau_1))) = K^-(\sigma'(f(\tau_1)))$ 
and $L^+(\sigma(\tau_1)) = L^+(\sigma'(\tau_1))$.  Again by 
\Eq{eq:inclusionk}, along with the property $f(\tau) > \tau$, 
we have therefore shown that
\be
  \sigma'(\tau_2) 
  \subset L^+(\sigma'(\tau_1)) \cap K^-(\sigma'(f(\tau_1))) 
  \subset L^+(\sigma'(\tau_1)) \cap K^-(\sigma'(\tau_1)) 
  = \mathring D(\Sigma^-(\sigma'(\tau_1)),
\ee
where $\Sigma(\sigma'(\tau_1))$ is a Cauchy surface split (into 
$\Sigma^\pm$) by $\sigma'(\tau_1)$.  It will be convenient to define 
the {\it outer wedge} $O_W(\sigma') = \mathring D[\Sigma^-(\sigma')]$. 
We thus find that every point in $\sigma'(\tau_2)$ is spacelike separated 
from every point in $\sigma'(\tau_1)$.  Hence, using this alternative 
definition of the generalized holographic screen, we automatically 
have that $H'$ is spacelike and directed outward when the corresponding 
portion of $H$ is spacelike.

Similarly, on the timelike part of $H$ (on which $\beta < 0$), 
we have $\sigma'(\tau_2) \subset N_{-l}(\sigma(\tau_2)) \subset 
\overline L^-(\sigma(\tau_2)) \subset L^{-}(\sigma(\tau_1))$ and 
further $\sigma'(\tau_2) \subset N_{+k}(\sigma(f(\tau_2))) \subset 
\overline K^-(\sigma(f(\tau_2))) \subset K^-(\sigma(f(\tau_1)))$, 
again by the condition $df/d\tau > 0$.  We further have in this 
case $K^-(\sigma(f(\tau_1))) = K^-(\sigma'(f(\tau_1)))$ and 
$L^-(\sigma(\tau_1)) = L^-(\sigma'(\tau_1))$.  Hence, again using 
\Eq{eq:inclusionk} and that $f(\tau) > \tau$, we have
\be
  \sigma'(\tau_2) 
  \subset L^-(\sigma'(\tau_1)) \cap K^-(\sigma'(f(\tau_1))) 
  \subset L^-(\sigma'(\tau_1)) \cap K^-(\sigma'(\tau_1)) 
  = I^{-}(\sigma'(\tau_1)),
\ee
so every point in $\sigma'(\tau_2)$ is in the chronological past of 
every point in $\sigma'(\tau_1)$.  Thus, we automatically have that 
$H'$ is timelike and past-directed when the corresponding portion of 
$H$ is timelike.

Therefore, with the simple requirements that $df/d\tau > 0$ and 
$f(\tau) \geq \tau$ (with equality when $H$ is null), we have an 
elegant construction of a generalized holographic screen $H'$, defined 
by its leaves as in \Eq{eq:leavesalt}, that automatically has the 
correct tangent and thus, by the argument in \Sec{subsec:area}, 
satisfies an area law.

\section{Outer Entropy}
\label{sec:outer}

Having established the general relativity results of \Sec{sec:GHS}, 
we now would like to understand their holographic interpretation.  In 
AdS/CFT~\cite{Maldacena:1997re,Gubser:1998bc,Witten:1998qj}, certain 
geometric quantities in the bulk have interpretations in terms of 
properties of the boundary CFT state.  The most celebrated example 
of this is the Ryu-Takayanagi relation~\cite{Ryu:2006bv,Ryu:2006ef} 
and its generalization to dynamical spacetimes by Hubeny, Rangamani, 
and Takayanagi~\cite{Hubeny:2007xt}, which relates the area of certain 
extremal surfaces in the bulk to the von~Neumann entropy
\be
  S[\rho] = -\mathrm{tr}\,\rho \log \rho
\ee
of the reduced density matrix $\rho$ on the homologous region on the 
boundary.  In particular, the HRT prescription implies that, for a 
boundary state $\rho$ corresponding to some classical bulk geometry 
with an extremal surface $X_{\rm HRT}$, the von~Neumann entropy 
satisfies
\be
  S[\rho] = \frac{A[X_{\rm HRT}]}{4G\hbar}.
\label{eq:HRT}
\ee
For a two-sided geometry in AdS/CFT described by a pure state, this 
entropy gives a measure of the entanglement between the boundary regions 
corresponding to the two sides of the spacetime split by the HRT surface. 
An extremal surface is defined to be a surface whose area is a local 
extremum as a functional over all surfaces in the bulk.  The HRT 
surface is chosen to be homologous to the boundary and an extremal 
surface of minimal area; such a surface can be identified using 
the maximin prescription~\cite{Wall:2012uf}. One can show that the 
HRT surface is a surface on which $\theta_k = \theta_l = 0$ and 
that there exists some Cauchy slice on which the area of the 
surface equals the minimal cross section of the slice.  While 
the HRT form of the entropy~\eqref{eq:HRT} has been extensively 
tested in AdS/CFT~\cite{Lewkowycz:2013nqa,Dong:2016hjy}, our results 
in this section will not need all of the structure of AdS/CFT for 
validity.  Instead, our conclusions will carry over under the 
assumption that the identification~\eqref{eq:HRT} can be made in 
any spacetime, that is, that there is some maximal extremal surface 
inside the black hole to which one can associate a fine-grained 
entropy for the ensemble.  This is the same set of assumptions used 
in \Ref{Engelhardt:2017aux}.  Moreover, if the holographic screen does 
indeed provide a boundary description of the spacetime in terms of 
a pure state, then this entropy would again equal the entanglement 
entropy between the boundary regions corresponding to the two sides 
of the spacetime split by the HRT surface.

There are compelling reasons why it is desirable to seek some entropic 
interpretation of the generalized holographic screens we considered in 
\Sec{sec:GHS}.  It has been conjectured that holographic screens play 
the role of the boundary of AdS in AdS/CFT for non-asymptotically-AdS 
spacetimes, enabling a suitable generalization of holography to arbitrary 
geometries~\cite{Nomura:2016ikr,Nomura:2017npr,Nomura:2017fyh,%
Nomura:2018kji}, although the details of this duality, including the 
explicit boundary theory, are not yet known.  If this is the case, then 
it is well motivated to ask whether there is a sense of renormalization 
in these holographic theories.  In AdS/CFT, renormalization group flow 
can be cast as motion in the bulk direction; formulating the theory on a 
surface at finite bulk coordinate yields a coarse-grained version of the 
original CFT~\cite{Akhmedov:1998vf,Alvarez:1998wr,Balasubramanian:1999jd,%
Skenderis:1999mm,deBoer:1999tgo}.  Thus, it is well motivated to ask 
whether the generalized holographic screens of \Sec{sec:GHS} play any 
similar coarse- or fine-grained role.  Indeed, one can view the area 
law discovered in \Sec{subsec:area} as evidence for some second law 
interpretation.

Furthermore, the fact that the event horizon itself is encompassed 
in the family of generalized holographic screens makes the quest 
for an entropic interpretation of these surfaces especially 
interesting.  The laws of black hole mechanics~\cite{Bekenstein:1972tm,%
Bekenstein:1973ur,Bardeen:1973gs,Hawking:1974rv,Hawking:1974sw} 
describing the dynamics of the event horizon ${\cal H}$ have direct 
thermodynamic interpretations, including Hawking's area theorem 
corresponding to the second law of thermodynamics and the 
Bekenstein-Hawking entropy,
\be
  S_{\rm BH} = \frac{A[{\cal H}]}{4G\hbar}.
\ee
Black hole thermodynamics was historically one of the original motivations 
for holography.  Despite this connection, however, there has previously 
been no direct interpretation of the event horizon itself from a 
holographic perspective.  Indeed, there are arguments showing that 
certain straightforward possibilities involving the area of the event 
horizon (i.e., the causal holographic information~\cite{Hubeny:2012wa}) 
cannot have a simple information-theoretic dual~\cite{Engelhardt:2017wgc}.

Previously, it was shown that the outermost spacelike portion of the 
holographic screen $H$ does possess a dual in terms of the von~Neumann 
entropy~\cite{Engelhardt:2017aux}.  Specifically, let us consider an 
outermost marginally-trapped surface (i.e., an {\it apparent horizon} 
$\sigma$), where by {\it outermost} we require that $\sigma$ is 
homologous to the boundary, with a partial spacelike Cauchy surface 
connecting $\sigma$ with the boundary such that any surface circumscribing 
$\sigma$ has area greater than that of $\sigma$.  Moreover, let us 
define the {\it outer entropy} associated with a codimension-two 
surface $\chi$,
\be
  \So{\chi} = \max_{\tilde \rho}(S[\tilde\rho]:\,O_W(\chi)),
\ee
as the entanglement entropy of one side of the entire boundary (computed 
via the HRT prescription) associated with the geometry described 
by the holographic state $\tilde \rho$, maximized over all possible 
$\tilde \rho$ corresponding to spacetimes $\tilde M$, satisfying 
the NCC, for which the {\it outer wedge} $O_W(\chi)$ is held fixed. 
In this sense, the outer entropy can be viewed as arising from 
the coarse-graining of the degrees of freedom associated with the 
(fine-grained) von~Neumann entropy; equivalently, it can be viewed 
as the maximum holographic entanglement entropy for one side of 
the boundary consistent with the outer wedge.  We recall from 
\Sec{subsec:intersection} that the outer wedge is defined as 
the set of points in $M$ spacelike separated from $\chi$ on the 
outer side, that is, $O_W(\chi) = \mathring D(\Sigma^-(\chi))$, 
where $\Sigma(\chi)$ is a Cauchy surface split by $\chi$.  With these 
definitions, the main result of \Ref{Engelhardt:2017aux} is that the 
outer entropy for the apparent horizon (the outermost spacelike part 
of the holographic screen) is given by its area:
\be
  \So{\sigma} = \frac{A[\sigma]}{4G\hbar}. 
\ee

We wish to relate the geometrical properties of leaves $\sigma'$ of the 
outermost spacelike or null part of the generalized holographic screen 
$H'$ defined in \Sec{sec:GHS} to their outer entropy.  We will show 
that $\So{\sigma'}$ is bounded from above by the area of $\sigma'$. 
Moreover, for the special case of spherically-symmetric spacetimes, 
we will provide an explicit formula for $\So{\sigma'}$ in terms of the 
geometry of $\sigma'$ (its area, curvature, etc.).

For the remainder of this paper, we will implicitly restrict ourselves 
to the outermost spacelike or null part of a generalized holographic 
screen $H'$, which we will write simply as $H'$.  That is, for any leaf 
$\sigma' \subset H'$ we consider in this and the following sections, 
we will take $\sigma'$ to be in $N_{-k}(\sigma)$ for $\sigma \subset H$ 
such that $\sigma$ is an outermost marginally-trapped surface in the 
sense of \Ref{Engelhardt:2017aux}.  Furthermore, in addition to the NEC, 
we will also impose the cosmological-constant-subtracted dominant energy 
condition ($\Lambda$DEC).  That is, writing the Einstein equation as
\be
  R_{ab} - \frac{1}{2}R\,g_{ab} + \Lambda\, g_{ab} = 8\pi G\,T_{ab},
\label{eq:Einstein}
\ee
we allow $\Lambda$ to take either sign but impose the dominant 
energy condition (DEC) on $T_{ab}$: $-T^a_{\;\;b}t^b$ is a causal, 
future-directed vector for all causal, future-directed vectors $t^a$.%
\footnote{Note that this is similar to, but somewhat stronger than, 
 the null dominant energy condition (NDEC), which requires the NEC 
 plus the stipulation that $-T^a_{\;\;b}k^b$ be a causal vector for 
 all null $k$.  While the NDEC allows for the cosmological constant 
 contribution, for either sign of $\Lambda$, to be folded into 
 $T_{ab}$, it does not bound the sign of $T_{kl}$ on its own, which 
 the $\Lambda$DEC does.}
This is essentially a causality requirement, enforcing that the positive 
flux of null energy not be superluminal as seen in any inertial frame. 
Finally, we will assume a generic condition on $\sigma'$, requiring 
that $\theta_k$ be strictly positive (rather than merely $\geq 0$) 
on $\sigma'$, so $\sigma' \not \subset H$, thus making the generalized 
holographic screen distinct from the original holographic screen. 
That is, the generalized holographic screen we consider in this and 
the next sections is foliated by leaves that are each {\it normal 
surfaces} (i.e., for which $\theta_l < 0$ and $\theta_k > 0$).

Let us first upper bound $\So{\sigma'}$ for some leaf $\sigma' \subset 
H'$.  We can choose the spacetime in the complement of $O_W(\sigma')$ 
to be the one that maximizes the area of the HRT surface $X_{\rm HRT}$. 
By definition, $\theta_k = \theta_l = 0$ on $X_{\rm HRT}$ and further 
there exists some Cauchy surface $\Sigma$ on which $X_{\rm HRT}$ is 
a surface of minimal cross-sectional area.  If $\sigma'\subset \Sigma$, 
we have $A[\sigma'] \geq A[X_{\rm HRT}]$ by definition of $\Sigma$. 
Moreover, if $\sigma'\subset I^+(\Sigma)$, then $N_{-k}(\sigma')$ 
intersects $\Sigma$ on some codimension-two surface $X^+$, while 
if $\sigma'\subset I^-(\Sigma)$, then $N_{+l}(\sigma')$ intersects 
$\Sigma$ on some codimension-two surface $X^-$.%
\footnote{An intersection of $N_l$ with $\Sigma$ is guaranteed 
 by a no-go theorem for topology change in general relativity: 
 since $M$ is by hypothesis globally hyperbolic, it has a Cauchy 
 surface and $M \simeq \Sigma \otimes \mathbb{R}$~\cite{Geroch:1970uw}, 
 so any causal hypersurface that completely divides the 
 spacetime---such as $N_l(\sigma')$ or $N_{-k}(\sigma') \cup 
 N_{+l}(\sigma')$~\cite{Akers:2017nrr}---must intersect any 
 Cauchy surface in a codimension-two surface of finite area.}
Since $X_{\rm HRT}$ is a surface of minimal cross-sectional area on 
$\Sigma$, it follows that $A[X_{\rm HRT}] \leq A[X^+]$ and $A[X_{\rm HRT}] 
\leq A[X^-]$.  By the Raychaudhuri equation~\eqref{eq:Raychaudhuri} 
in the $k$ direction and the fact that $\theta_k > 0$ on $\sigma'$, 
it follows that $\theta_k > 0$ on the entire segment of $N_{-k}(\sigma')$ 
between $X^+$ and $\sigma'$, so $A[X^+] < A[\sigma']$.  Similarly, the 
Raychaudhuri equation in the $l$ direction is
\be
  \nabla_l \theta_l = -\frac{1}{D-2}\theta_l^2 
    - \varsigma_l^2 - R_{ab} l^a l^b,
\label{eq:Raychaudhuril}
\ee
where $\varsigma_l$ is the shear of the $l$ congruence and $\nabla_l 
= l^a \nabla_a$.  As a result, since $\theta_l < 0$ on $\sigma'$, we 
have $\theta_l < 0$ on the entire segment of $N_{+l}(\sigma')$ between 
$\sigma'$ and $X^-$, so $A[X^-] < A[\sigma']$.  Since we have been 
considering the spacetime in which the area of $X_{\rm HRT}$ is maximal 
for fixed $O_W(\sigma')$, we have $\So{\sigma'} = A[X_{\rm HRT}]/4G\hbar$. 
We thus obtain an upper bound on the outer entropy of $\sigma'$:%
\footnote{\Ref{Engelhardt:2018kcs} reaches a similar conclusion.}
\be
  \So{\sigma'} \leq \frac{A[\sigma']}{4G\hbar}.
\label{eq:upper}
\ee

\section{Holographic Dual for Spherically-Symmetric Spacetimes}
\label{sec:sphere}

Beyond the upper bound in \Eq{eq:upper}, we would like to have an 
explicit expression for the outer entropy $\So{\sigma'}$, defined 
in \Sec{sec:outer}, for the generalized holographic screen constructed 
in \Sec{sec:GHS}.  While there are subtleties for general spacetimes, 
we can derive an explicit expression in the case of spherically-symmetric 
surfaces $\sigma'$.

Before assuming spherical symmetry, let us first establish some 
intermediate results.  First, we note that, for the $\sigma \subset H$ 
for which $\sigma'\subset N_{-k}(\sigma)$, there exists (since by 
hypothesis $\sigma$ is an outermost marginally-trapped surface) a 
partial Cauchy surface $\Sigma \subset O_W(\sigma)$ such that for 
any slice $\rho$ of $\Sigma$, which by definition subtends $\sigma$, 
$A[\rho]>A[\sigma]$.  Such a partial Cauchy surface also exists for 
$\sigma'$, since $\theta_k \geq 0$ between $\sigma'$ and $\sigma$ 
and is positive at $\sigma'$: simply take the union of $ \Sigma$ and 
$N_{-k}(\sigma) \cap N_{+k}(\sigma')$.  Thus, there exists a Cauchy 
surface $\Sigma' \supset \sigma'$ for which $\Sigma'^-$ connects 
$\sigma'$ with the boundary and such that every slice $\rho \subset 
\Sigma'^-$ satisfies $A[\rho] > A[\sigma']$.

We can prove that $X_{\rm HRT}$ is in $\overline D(\Sigma'^+)$, the 
closure of the domain of dependence of $\Sigma'^+$, the interior partial 
Cauchy surface ending on $\sigma'$.  We recall that $I^\pm(\sigma')$ and 
$\overline D(\Sigma'^\pm) - \sigma'$ form a partition of $M - \sigma'$. 
Suppose that $X_{\rm HRT} \not \subset \overline D(\Sigma'^+)$. 
Then either $N_{-l}(X_{\rm HRT})$ or $N_{+k}(X_{\rm HRT})$ intersects 
$\Sigma'^-$ on some surface $\zeta$.  We have $A[\zeta] > A[\sigma']$. 
Moreover, by the Raychaudhuri equation along $N_{-l}(X_{\rm HRT})$ 
and $N_{+k}(X_{\rm HRT})$, we have $A[X_{\rm HRT}] \geq A[\zeta]$. 
Hence, $A[X_{\rm HRT}] > A[\sigma']$, in contradiction with the result 
established in \Sec{sec:outer} that $A[\sigma'] \geq A[X_{\rm HRT}]$. 
We therefore must have $X_{\rm HRT} \subset \overline D(\Sigma'^+)$.

\subsection{Construction}
\label{subsec:construction}

In order to place a lower bound on $\So{\sigma'}$ for a 
spherically-symmetric $\sigma'$, it suffices to analyze spacetimes 
that are also spherically symmetric in the interior of $\sigma'$; 
for these geometries, we can find the maximal HRT surface and calculate 
its area.  We will do this presently and subsequently argue that our 
construction is optimal over {\it all} geometries, producing the HRT 
surface of maximal area for fixed $O_W(\sigma')$, so our lower bound 
is in fact saturated.

To construct our spacetime outside of $O_W(\sigma')$, we will use 
the {\it characteristic initial data formalism}~\cite{Rendall:2000,%
Brady:1995na,ChoquetBruhat:2010ih,Luk:2011vf,Chrusciel:2012ap,%
Chrusciel:2012xf,Chrusciel:2014lha} as in \Ref{Engelhardt:2017aux}. 
Given a Cauchy surface formed by light sheets, the characteristic initial 
data formalism implies that a spacetime exists for self-consistent 
initial data satisfying the {\it constraint equations}.  For the null 
portion of a Cauchy surface in the $k$ direction, the constraint equations 
are~\cite{Price:1986yy,Hayward:1993wb,Hayward:2004fz,Gourgoulhon:2005ng,%
Hayward:2006ss,Cao:2010vj,Sousa:2017auc}
\be
\begin{aligned}
  \nabla_k \theta_k &= -\frac{1}{D-2}\theta_k^2 - \varsigma_k^2 - G_{kk} 
    & \hspace{3cm}\text{[Raychaudhuri]}
\\
  {\cal L}_k \omega_i &= -\theta_k \omega_i + \frac{D-3}{D-2}{\cal D}_i 
      \theta_k - ({\cal D} \cdot \varsigma_k)_i + G_{ik} 
    & \text{[Damour-Navier-Stokes]}
\\
  \nabla_k \theta_l &= -\frac{1}{2}{\cal R} - \theta_k \theta_l 
      + \omega^2 + {\cal D}\cdot\omega + G_{kl}, 
    & \text{[Cross-focusing]}
\label{eq:constraintk}
\end{aligned}
\ee
while for a null portion of a Cauchy surface in the $l$ direction, the 
constraint equations become
\be
\begin{aligned}
  \nabla_l \theta_l &=  -\frac{1}{D-2}\theta_l^2 - \varsigma_l^2 - G_{ll} 
    &\hspace{3cm}\text{[Raychaudhuri]}
\\
  {\cal L}_l \omega_i &= -\theta_l \omega_i - \frac{D-3}{D-2}{\cal D}_i 
      \theta_l + ({\cal D} \cdot \varsigma_l)_i - G_{il} 
    & \text{[Damour-Navier-Stokes]}
\\
  \nabla_l \theta_k &= -\frac{1}{2}{\cal R} - \theta_k \theta_l 
      + \omega^2 - {\cal D}\cdot\omega + G_{kl}. 
    & \text{[Cross-focusing]}
\label{eq:constraintl}
\end{aligned}
\ee
Here, $\cal R$ is the intrinsic Ricci curvature of the codimension-two 
slices at constant affine parameter and $G_{ab}$ is the Einstein tensor, 
$R_{ab} - \frac{1}{2} R g_{ab}$.  The twist one-form gauge field (the 
H{\'a}{\'\j}i{\v c}ek one-form) is $\omega_i = \frac{1}{2} q_{ib} 
{\cal L}_k l^b$, where $q_{ab} = g_{ab} + k_a l_b + l_a k_b$ is 
the induced metric.  Lie derivatives are denoted by ${\cal L}$, 
while ${\cal D}$ is the transverse covariant derivative within the 
codimension-two surface.  We use letters $a,b$ for $D$-dimensional 
spacetime indices, $i,j$ for $(D-2)$-dimensional transverse spatial 
indices in the codimension-two surface, and indices $k$ and $l$ for 
a $D$-dimensional spacetime index contracted into null vectors $k^a$ 
and $l^a$, respectively.

The junction conditions mandate continuity of $\theta_k$, $\theta_l$, 
and $\omega_i$, while $\varsigma_k$ and $\varsigma_l$ can change 
discontinuously via an appropriate shock wave in the Weyl 
tensor~\cite{Wald} (i.e., gravitational waves~\cite{Luk:2012hi,%
Luk:2013zr}).  We choose $k$ and $l$ to be affinely parameterized 
tangent vectors to null geodesic congruences originating orthogonally 
from the codimension-two surfaces we consider.  We further specify 
the relative normalization of these vectors to be $k \cdot l = -1$, 
so $g_{kl} = -1$.  These choices eliminate other terms that 
could have appeared in the Damour-Navier-Stokes equations in 
\Eqs{eq:constraintk}{eq:constraintl}~\cite{Hayward:2006ss}.  On the 
$k$ and $l$ congruences, we can define affine parameters $\nu$ and 
$\mu$, respectively, normalized such that $\nabla_\nu = \nabla_k$ and 
$\nabla_\mu = \nabla_l$.  Using the Einstein equation~\eqref{eq:Einstein}, 
we can replace $G_{kl}$ by $8\pi G\, T_{kl} - \Lambda g_{kl} = 8\pi G\, 
T_{kl} + \Lambda$, $G_{kk}$ by $8\pi G\,T_{kk}$, and $G_{ll}$ by 
$8\pi G\,T_{ll}$.  The transverse coordinates $x^i$ are chosen 
to always lie within the codimension-two surface of constant 
affine parameter.

For now, we restrict to a spherically-symmetric spacetime in the 
interior of $\sigma'$, i.e., in $\overline D(\Sigma'^+)$.  Requiring 
the energy-momentum tensor to respect the $SO(D-1)$ invariance of 
spherical symmetry, we must have $T_{ik} = T_{il} = 0$.  Similarly, 
the shears $\varsigma_k$ and $\varsigma_l$ both vanish, as does the 
twist one-form $\omega_i$.  Hence, for spherical spacetimes satisfying 
the Einstein equation, the constraint equations \eqref{eq:constraintk} 
and \eqref{eq:constraintl} become
\be
\begin{aligned}
  \nabla_k \theta_k &= -\frac{1}{D-2}\theta_k^2 - 8\pi G\,T_{kk} 
  \qquad & \hspace{1cm}\text{[Raychaudhuri]}
\\
  \nabla_k \theta_l &= -\frac{1}{2}{\cal R} - \theta_l\theta_k 
    + 8\pi G\,T_{kl} + \Lambda 
  \qquad & \text{[Cross-focusing]}
\label{eq:constraintksphere}
\end{aligned}
\ee
and
\be
\begin{aligned}
  \nabla_l \theta_l &= -\frac{1}{D-2}\theta_l^2 - 8\pi G\,T_{ll} 
  \qquad & \hspace{1cm}\text{[Raychaudhuri]}
\\
  \nabla_l \theta_k &= -\frac{1}{2}{\cal R} - \theta_l\theta_k 
    + 8 \pi G\,T_{kl} + \Lambda. 
  \qquad & \text{[Cross-focusing]}
\label{eq:constraintlsphere}
\end{aligned}
\ee
While the NEC requires that $T_{kk}$ and $T_{ll}$ be nonnegative, the 
$\Lambda$DEC imposes similar condition on $T_{kl}$.  We can rewrite the 
$\Lambda$DEC as the requirement that $T_{ab} t_1^a t_2^b \geq 0$ for all 
causal, future-directed vectors $t_1$ and $t_2$.  Making the particular 
choice $t_1 = k$ and $t_2 = l$, we have $T_{kl} \geq 0$.

For a spherically-symmetric spacetime, there is a nice relation between 
the intrinsic Ricci curvature $\cal R$ and the null expansion.  For a 
$(D-2)$-sphere of radial coordinate $r$,
\be
  {\cal R} = \frac{(D-2)(D-3)}{r^{2}},
\label{eq:Rdefsph}
\ee
which implies $\nabla_{k}(\log{\cal R}) = -(2/r)(dr/d\nu)$.  Writing 
$A \propto r^{D-2}$ for the area of the constant-$\nu$ cross section 
of $N_{-k}(\sigma')$, we therefore have
\be
  \theta_{k} = \frac{\nabla_{k}A}{A} 
  = \frac{D-2}{r} \frac{dr}{d\nu} 
  = -\frac{D-2}{2}\nabla_{k}\log{\cal R}.
\label{eq:Rconst}
\ee

Given $O_W(\sigma')$, let us now construct a particular spacetime and 
compute its HRT surface.  On $N_{-k}(\sigma')$, we will choose data 
with $T_{kk} = 0$.  Hence, we can solve the Raychaudhuri equation in 
\Eq{eq:constraintksphere} to compute $\theta_k(\nu)$ on $N_{-k}(\sigma')$:
\be
  \theta_{k}(\nu) 
  = \left[\frac{1}{\theta_{k}[\sigma']} + \frac{\nu}{D-2}\right]^{-1},
\label{eq:thetaksoln}
\ee
where we define $\sigma'$ to correspond to the $\nu = 0$ surface.  Thus, 
$N_{-k}(\sigma')$ encounters a caustic at affine parameter
\be
  \nu_{\rm c} = -\frac{D-2}{\theta_{k}[\sigma']}.
\label{eq:nuc}
\ee
Using the relation \eqref{eq:Rconst}, we have
\be
  {\cal R}(\nu) = {\cal R}[\sigma'] 
    \exp\biggl[-\frac{2}{D-2}\int_0^\nu\!\theta_k(\nu)\,d\nu\biggr] 
  = \frac{{\cal R}[\sigma']}
    {\left[1+\frac{\theta_k[\sigma']\,\nu}{D-2}\right]^2} 
  = \left[\frac{\theta_k(\nu)}{\theta_k[\sigma']}\right]^2 
    {\cal R}[\sigma'].
\label{eq:Rsoln}
\ee
Note that if ${\cal R}(\nu)$ and $\theta_{k}(\nu)$ diverge to $+\infty$, 
they do so together, as $r\rightarrow 0$.  However, there exist spacetimes 
that do not have $r\rightarrow 0$ accessible along $N_{-k}(\sigma')$ and 
hence do not possess a caustic.

Let us define a surface $X \subset N_{-k}(\sigma')$ on which $\theta_l 
= 0$.  For $X$ to exist, we must choose our data on $N_{-k}(\sigma')$ 
such that the affine parameter $\nu_0$ on which $\theta_l$ vanishes 
satisfies $\nu_0 > \nu_{\rm c}$.  We choose $T_{kl}$ to vanish on 
$N_{-k}(\sigma')$.%
\footnote{We can make the choice of $T_{kl}$ and $T_{kk}$ vanishing 
 on $N_{-k}(\sigma')$ consistently with energy-momentum conservation 
 $\nabla^a T_{ab} = 0$, the NEC, the $\Lambda$DEC, and smoothness via 
 a regularization procedure, in which we consider a shell of matter 
 occupying a thin slice of $N_{-k}(\sigma')$ adjacent to $\sigma'$, 
 then take the limit as the shell thickness goes to zero.}
Without loss of generality, let us write $\theta_l(\nu)$ on 
$N_{-k}(\sigma')$ as
\be
  \theta_l(\nu) 
  = \frac{\theta_k[\sigma'] \theta_l[\sigma'] q(\nu)}{\theta_k(\nu)}
\label{eq:qdef}
\ee
for some function $q(\nu)$ that satisfies $q(\nu=0) = 1$ and 
$q(\nu=\nu_0) = 0$ on $X$, that is, for some $\nu_0 \in (\nu_{\rm c},0)$. 
Since we seek the first time $\theta_l$ vanishes when going from $\sigma'$ 
along the $-k$ congruence, without loss of generality we can take 
$q(\nu) > 0$ for $\nu \in (\nu_0,0]$.  The cross-focusing equation 
in \Eq{eq:constraintksphere}, combined with \Eqs{eq:thetaksoln}{eq:Rsoln}, 
then becomes
\be
  (a\nu+b)^3 q' + (a\nu+b)^2 (cq + d) = e,
\ee
where the constants $a,b,c,d,e$ are given by
\be 
\begin{aligned}
 & a = \frac{1}{D-2},&&\qquad
   b = \frac{1}{\theta_{k}[\sigma']},&\qquad
   c = \frac{D-1}{D-2},
\\
 & d = -\frac{\Lambda}{\theta_k[\sigma']\theta_l[\sigma']},&&\qquad
   e = -\frac{{\cal R}[\sigma']}
        {2(\theta_{k}[\sigma'])^{3}\theta_{l}[\sigma']}.
\end{aligned}
\ee
The general solution is 
\be
  q(\nu) = \frac{e}{(c-2a)(a\nu+b)^2} 
    + m(a\nu+b)^{-\frac{c}{a}} - \frac{d}{c},
\ee
where $m$ is a constant of integration that we fix by demanding 
$q(\nu=0) = 1$.  That is,
\be
\begin{aligned}
  q(\nu) &= \left[1+\frac{\theta_{k}[\sigma']\nu}{D-2}\right]^{-(D-1)} 
    + \frac{D-2}{D-1} \frac{\Lambda}{\theta_k[\sigma']\theta_l[\sigma']} 
    \left\{ 1 - \left[1+\frac{\theta_k[\sigma']\nu}{D-2}\right]^{-(D-1)} 
    \right\}
\\
 & \qquad +\frac{1}{2}\frac{D-2}{D-3} 
    \frac{{\cal R}[\sigma']}{\theta_k[\sigma']\theta_l[\sigma']} 
    \left\{ \left[1+\frac{\theta_k[\sigma']\nu}{D-2}\right]^{-(D-1)} 
    - \left[1+\frac{\theta_k[\sigma']\nu}{D-2}\right]^{-2} \right\}.
\end{aligned}
\ee
Defining
\be
\begin{aligned}
  \xi(\nu) &= \frac{\theta_k(\nu)}{\theta_k[\sigma']},
\\
  \rho &= -\frac{1}{2}\frac{D-2}{D-3} 
    \frac{{\cal R}[\sigma']}{\theta_k[\sigma']\theta_l[\sigma']},
\\
  \lambda &= \frac{D-2}{D-1} 
    \frac{\Lambda}{\theta_k[\sigma']\theta_l[\sigma']},
\label{eq:rholambdadef}
\end{aligned}
\ee
we can rewrite $q$ simply as
\be
  q(\nu) = (1-\rho-\lambda)\xi(\nu)^{D-1} + \rho\,\xi(\nu)^2 + \lambda.
\label{eq:ffinal}
\ee

\begin{figure}[t]
\centering
  \includegraphics[width = 8.5cm]{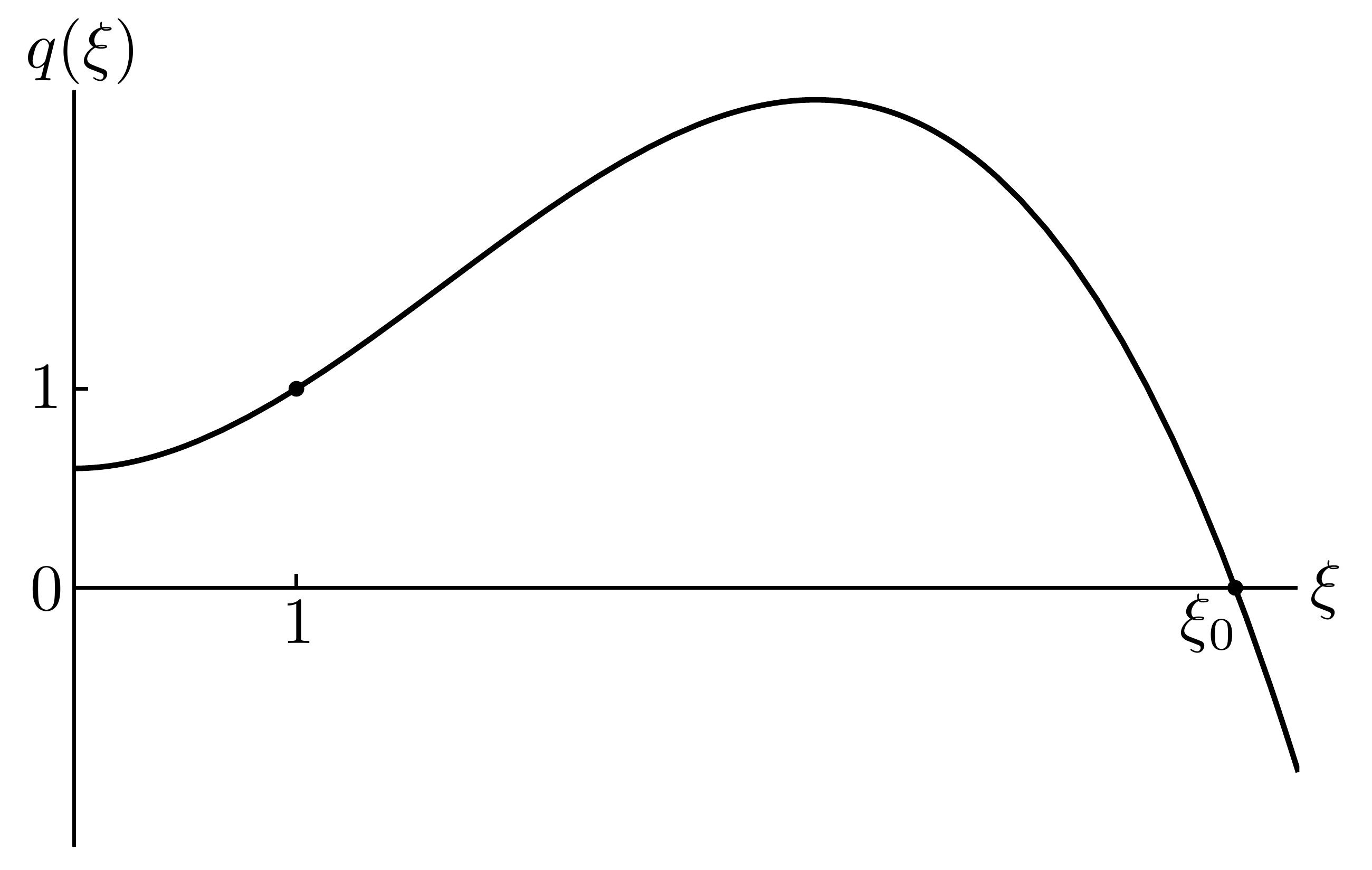}
\caption{Example of the polynomial $q(\xi) = (1-\rho-\lambda)\xi^{D-1} 
 + \rho \xi^2 + \lambda$.  By definition, $q(\xi=1) = 1$ and $\rho > 0$, 
 but $\lambda$ is allowed to take either sign.  For $D \geq 4$, 
 there is exactly one real zero at $\xi_0 > 1$ if and only if 
 $\rho + \lambda > 1$.  For small $\xi$, the polynomial behaves 
 like $\rho\xi^2 + \lambda$, while for large $\xi$, the dominant 
 contribution is $(1-\rho-\lambda)\xi^{D-1}$.}
\label{fig:q}
\end{figure}

By definition, $\rho > 0$.  For now, we will take $D \geq 4$, postponing 
a discussion of the special case of $D=3$ to \Sec{subsubsec:3D}.  The 
polynomial in \Eq{eq:ffinal} will have a single zero at some real value 
of $\xi = \xi_0 > 1$ if and only if
\be
  \rho + \lambda > 1.
\label{eq:rhopluslambda}
\ee
See \Fig{fig:q} for an illustration of $q$ as a polynomial in $\xi$. 
This zero corresponds to the surface $X$ on which $\theta_l = 0$, at 
affine parameter
\be
  \nu_0 = \frac{D-2}{\theta_k[\sigma']}\left(\frac{1}{\xi_0} - 1\right).
\ee
Since by assumption $\xi_0 \in (1,\infty)$, we have $\nu_0 \in 
(\nu_{\rm c},0)$, so $X$ indeed exists with $\theta_k$ having no caustic 
along $N_{+k}(X) \cap N_{-k}(\sigma') = \Sigma_1$.  The area of $X$ is
\be
  A[X] = A[\sigma'] \exp\left[\int_{0}^{\nu_0}\theta_k(\nu)d\nu\right] 
  = \frac{A[\sigma']}{\xi_0^{D-2}}.
\label{eq:areaX}
\ee
For general $\rho$, $\lambda$, and $D$, there is no closed-form 
expression for the zero of \Eq{eq:ffinal}, even if it exists.  For 
the present, we will continue to write the zero as $\xi_0$ and will 
later consider the cases in which either $\rho$ or $\lambda$ is 
negligible, allowing the zero to be analytically expressed.

From $X$, we will follow $N_{+l}(X)$, holding $\theta_l = 0$ fixed, 
so that the area is stationary along the light sheet.  This requires 
setting $T_{ll} = 0$ to satisfy the Raychaudhuri equation in 
\Eq{eq:constraintlsphere}.  We also set $T_{kl} = 0$ and hold 
${\cal R}$ fixed.

Consider the polynomial $q(\xi) = (1-\rho-\lambda)\xi^{D-1} + \rho\xi^2 
+ \lambda$.  From the fact that $q(\xi = 1) = 1$, that $\xi_0$ gives 
the unique real zero of $q(\xi)$ for $\xi_0 > 1$, and that $q(\xi) < 0$ 
for sufficiently large $\xi$, we must have  $dq/d\xi < 0$ at $\xi = 
\xi_0$.  By \Eq{eq:ffinal}, this requirement implies
\be
  (D-3) \rho \xi_0^2 + (D-1) \lambda > 0.
\label{eq:rholambdacond2}
\ee
Using \Eq{eq:rholambdacond2}, along with the definition of $\xi$ 
in \Eq{eq:rholambdadef} and its relation to ${\cal R}(\nu)$ in 
\Eq{eq:Rsoln}, we therefore have
\be 
\begin{aligned}
  -\frac{1}{2}{\cal R}[X] + \Lambda 
  &= -\frac{1}{2} \xi_0^2 {\cal R}[\sigma'] + \Lambda 
  = \frac{\theta_k[\sigma']\theta_l[\sigma']}{D-2} 
    [(D-3)\rho\xi_0^2 + (D-1)\lambda] 
  < 0.
\end{aligned}
\ee
Hence, from the cross-focusing equation in \Eq{eq:constraintlsphere}, 
we find that $\nabla_l \theta_k < 0$ on $N_{+l}[X]$, so there will be 
some value $\mu_0$ of the affine parameter $\mu$ for which $\theta_k$ 
vanishes.  The surface $\tilde X$ at $\mu = \mu_0$ satisfies $\theta_k 
= \theta_l = 0$.

\begin{figure}[t]
\centering
  \includegraphics[width = 11cm]{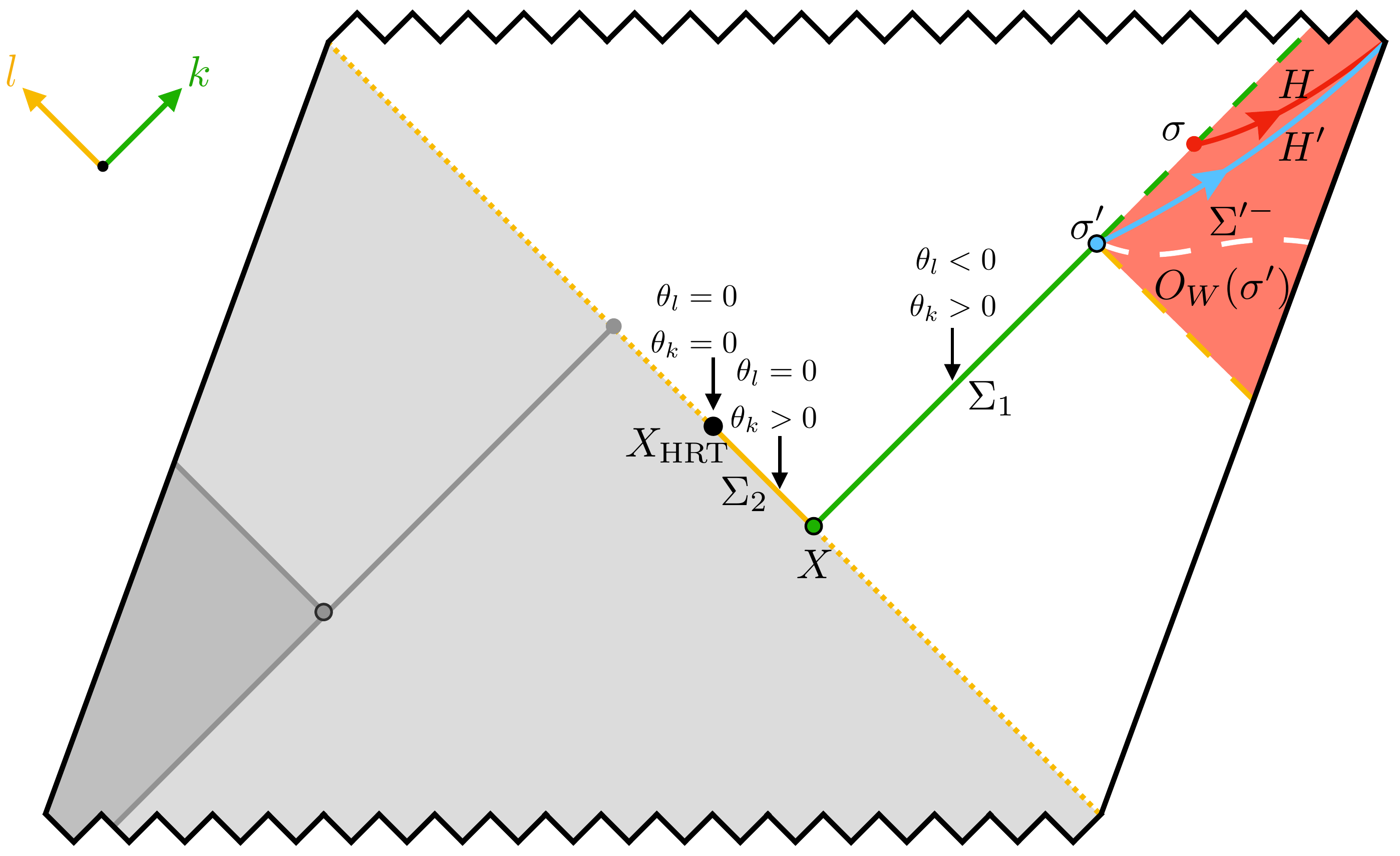}
\caption{Illustration of the construction of the HRT surface using the 
 characteristic initial data formalism.  The outer wedge $O_W(\sigma')$ 
 (red shading) of $\sigma'$ (blue dot) is held fixed.  We flow along 
 $N_{-k}(\sigma')$ until we reach a marginally antitrapped surface $X$. 
 We then flow along $N_{+l}(X)$ while keeping cross sections of the 
 light sheet stationary, until we reach a surface $X_{\rm HRT}$ where 
 $\theta_k = \theta_l = 0$ (black dot).  The spacetime is completed 
 (gray shading) by CPT reflection across $N_l(X_{\rm HRT})$ (orange 
 solid and dotted lines).  The partial Cauchy surface $\Sigma'^-$ (white 
 dashed line) connecting $\sigma'$ with the boundary by hypothesis 
 satisfies $A[\rho] > A[\sigma']$ for all cross sections $\rho \subset 
 \Sigma'^-$.  We note that $X_{\rm HRT}$ has minimal cross-sectional 
 area on the Cauchy slice formed by the union of $\Sigma'^-$, 
 $\Sigma_1 = N_{-k}(\sigma') \cap N_{+k}(X)$ (green solid line), 
 and $\Sigma_2 = N_{+l}(X) \cap N_{-l}(X_{\rm HRT})$ (orange solid 
 line), along with their CPT reflections, so $X_{\rm HRT}$ is indeed 
 an HRT surface.}
\label{fig:const}
\end{figure}

We can complete the entire spacetime by CPT reflection about $\tilde X$. 
Furthermore, defining $\Sigma_2 = N_{-l}(\tilde X) \cap N_{+l}(X)$, we 
observe that $\tilde X$ is a minimal cross section on the Cauchy slice 
$\tilde \Sigma$ formed by $\Sigma'^- \cup \Sigma_1 \cup \Sigma_2$ and 
its CPT reflection.  As a result, any other extremal surface $\hat X$ 
will have greater area than $\tilde X$, following the argument in 
\Ref{Engelhardt:2017aux}: by the Raychaudhuri equation any slice 
of $N_k(\hat X)$ has area upper bounded by that of $\hat X$ and 
furthermore the intersection of $N_k(\hat X)$ with $\tilde \Sigma$ 
will have area lower bounded by that of $\tilde X$, so $A[\tilde X] 
\leq A[\hat X]$.  Hence, $\tilde X$ is an HRT surface, which we will 
henceforth label as $X_{\rm HRT}$.  The area of $X_{\rm HRT}$ equals 
$A[X]$ by construction.  We have thus constructed a lower bound for 
$\So{\sigma'}$:
\be
  \So{\sigma'} \geq \frac{A[X_{\rm HRT}]}{4G\hbar} 
  = \frac{A[X]}{4G\hbar} = \frac{A[\sigma']}{4G\hbar \xi_0^{D-2}}.
\label{eq:lower}
\ee
Our construction is summarized in \Fig{fig:const}.

\subsection{Optimization}
\label{subsec:opt}

We now argue that our construction in \Sec{subsec:construction} is in 
fact optimal.  Namely, for a spherically-symmetric $\sigma'$ with its 
outer wedge fixed, the construction produces the spacetime that has 
the HRT surface with the largest possible area (subject to the NEC 
and $\Lambda$DEC).  This implies that our lower bound in \Eq{eq:lower} 
is actually an equality.

\begin{figure}[t]
\centering
  \includegraphics[width = 8cm]{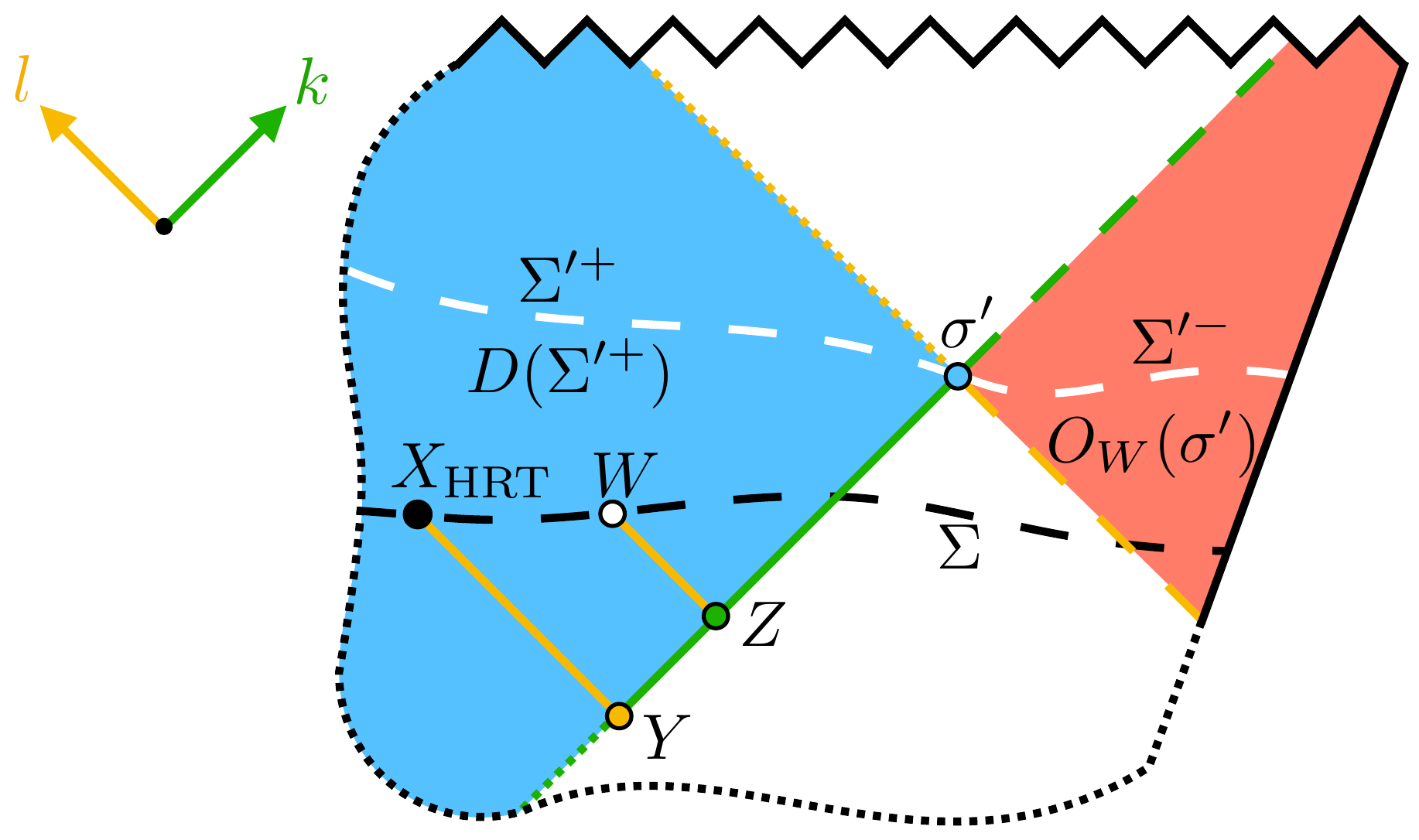}
\caption{Illustration of various definitions appearing in the procedure 
 for maximizing the area of the HRT surface while keeping the outer 
 wedge $O_W(\sigma')$ (red shading) of $\sigma'$ (blue dot) held fixed. 
 The HRT surface $X_{\rm HRT}$ (black dot) must appear in the closure 
 of the inner domain of dependence $D(\Sigma'^+)$ (blue shading) of 
 a Cauchy surface $\Sigma'$ passing through $\sigma'$ (white dashed 
 line), so the surface $Y = N_{-l}(X_{\rm HRT}) \cap N_{-k}(\sigma')$ 
 exists, on which $\theta_l \geq 0$.  By continuity, there must 
 exist a surface $Z\subset N_{+k}(Y) \cap N_{-k}(\sigma')$ on which 
 $\theta_l = 0$.  By definition, there exists a Cauchy surface 
 $\Sigma \supset X_{\rm HRT}$ for which $X_{\rm HRT}$ has the minimal 
 cross-sectional area.  Since $A[Z] \geq A[W]$, where $W = N_l(Z) 
 \cap \Sigma$, it follows that $A[Z]\geq A[X_{\rm HRT}]$.}
\label{fig:opt}
\end{figure}

We begin by considering an {\it arbitrary} spacetime satisfying our 
energy conditions and with the outer wedge of $\sigma'$ fixed.  Since 
$X_{\rm HRT} \subset \overline D(\Sigma'^+)$, $N_{-l}(X_{\rm HRT}) 
\cap N_{-k}(\sigma')$ is nonempty and, in particular, is some 
codimension-two surface $Y$; see \Fig{fig:opt}.  Now, $A[Y] \leq 
A[\sigma']$, since $\theta_k > 0$ along $N_{-k}(\sigma')$.  The fact 
that $\theta_l = 0$ on $X_{\rm HRT}$ implies $\theta_l[Y] \geq 0$, 
so since $\theta_l[\sigma'] < 0$ by construction, $Y \neq \sigma'$ 
and $A[Y] < A[\sigma']$.  By continuity, there must be some surface 
$Z \subset N_{+k}(Y) \cap N_{-k}(\sigma')$ for which $\theta_l[Z] = 0$. 
We have $A[Z] < A[\sigma']$ and, if $Z \neq Y$, $A[Z] > A[Y]$.

Recalling the definition of $\Sigma$ as a Cauchy surface on which 
$X_{\rm HRT}$ has minimal cross-sectional area, we can define the 
codimension-two surface $W = N_l(Z) \cap \Sigma$, which by definition 
satisfies $A[W] \geq A[X_{\rm HRT}]$.  Since $\theta_l = 0$ on $Z$, 
it follows from the Raychaudhuri equation \eqref{eq:Raychaudhuril} 
that slices of $N_l(Z)$ have areas upper bounded by $A[Z]$, so 
$A[W] \leq A[Z]$ and hence $A[Z] \geq A[X_{\rm HRT}]$.

To compute $\So{\sigma'}$, we must maximize the area of the HRT surface 
or, equivalently, minimize the quantity
\be
  \Delta A = A[\sigma'] - A[X_{\rm HRT}]
\ee
over all spacetimes with the geometry of $\sigma'$ held fixed.  Let us 
write $\Delta A$ as the sum of $\Delta A_1$ and $\Delta A_2$, where
\be
\begin{aligned}
  \Delta A_1 &= A[\sigma'] - A[Z]
\\
  \Delta A_2 &= A[Z]-A[X_{\rm HRT}].
\end{aligned}
\ee
We note that $\Delta A_1 > 0$ and $\Delta A_2 \geq 0$.  A sufficient 
condition for minimizing $\Delta A$ is to simultaneously minimize 
$\Delta A_1$ and $\Delta A_2$.

While we have taken $\sigma'$ to be spherically symmetric, the quantity 
$\So{\sigma'}$ is in general maximized over all possible spacetimes 
with $O_W(\sigma')$ held fixed; in particular, $N_{-k}(\sigma')$ could 
a~priori break spherical symmetry.  Even if this happens, we would take 
the affine parameter $\nu$, which is now defined separately for each 
generator of $N_{-k}(\sigma')$, to respect spherical symmetry at 
$\sigma'$.  Specifically, we choose $\nu = 0$ at $\sigma'$ and take 
the normalization of $\nu$ such that $\theta_k$ is uniform over 
$\sigma'$.

Let us first choose the data on $N_{-k}(\sigma') \cap N_{+k}(Z)$ 
to minimize $\Delta A_1$.  Because of the Raychaudhuri equation 
\eqref{eq:Raychaudhuri} and the NEC, a given area element $\delta A$ 
can only decrease toward the $-k$ direction (recalling that $\theta_{k} 
= d\log\delta A/d\nu$).  Hence, we optimize the area of $Z$ by 
taking $\varsigma_k = T_{kk} = 0$ along each null geodesic generating 
$N_{-k}(\sigma')$.  This implies that without a~priori assuming 
spherical symmetry, we have deduced that the area elements at surfaces 
of constant $\nu$ are maximized if they are all given by a simple 
rescaling of the original area element:
\be
  \delta A(\nu) 
  = \delta A[\sigma'] \exp\left[\int_{0}^{\nu}\theta_k(\nu)d\nu\right] 
  = \delta A[\sigma'] 
    \left[1+\frac{\nu\theta_{k}[\sigma']}{D-2}\right]^{D-2}.
\label{eq:simple-resc}
\ee
In particular, the metric on a constant-$\nu$ surface is given simply 
by conformally rescaling that on $\sigma'$, so it is spherically 
symmetric.  Hence, our optimization of $\Delta A_1$ implies, given 
a spherically-symmetric surface $\sigma'$, that $N_{-k}(\sigma')$ 
is also spherical on surfaces of constant affine parameter.%
\footnote{This conclusion is closely related to the light-cone 
 theorem~\cite{ChoquetBruhat:2009fy}, which uses stronger assumptions 
 about the energy conditions but a more general geometric setup.}

Since we now know that the geometry on $N_{-k}(\sigma')$ respects 
spherical symmetry, we expect to have $\omega_i = T_{ik} = 0$ there 
as well.  This conclusion can also be understood as a consequence of 
the $\Lambda$DEC and our choice of $T_{kk} = 0$, via the following 
argument.  By the $\Lambda$DEC, $-T^a_{\;\;b}t^b$ is a causal vector 
for all causal $t$, so in particular $v^a = -T^a_{\;\;k}$ is causal. 
By choosing $T_{kk} = 0$ along $N_{-k}(\sigma')$, we have $v \cdot k 
= 0$, so $v\propto k$.  Since the transverse coordinates are by 
definition orthogonal to $k$, $v_i$ vanishes, so $T_{ik}=0$; see 
\Ref{ChoquetBruhat:2009fy}.  By our choice $T_{kk} = \varsigma_k = 0$, 
the Raychaudhuri equation implies that ${\cal D}_i \theta_k = 0$, 
as seen in \Eq{eq:simple-resc}.  Hence, the Damour-Navier-Stokes 
equation in \Eq{eq:constraintk} becomes simply ${\cal L}_k \omega_i 
= -\theta_k \omega_i$, which, given the initial condition that 
$\omega_i[\sigma'] = 0$ (by spherical symmetry of $\sigma'$), implies 
that $\omega_i = 0$ along the entirety of $N_{-k}(\sigma')$ as expected.

The above choice of the data, $\varsigma_k = T_{kk} = 0$, only minimizes 
$\nabla_k \theta_k$. To actually minimize $\Delta A_1$, we must also
make $\nabla_k \theta_l$ as large and negative as possible, in order
to bring the $\theta_l = 0$ surface, $Z$, to its minimum affine distance 
from $\sigma'$; see \Eq{eq:simple-resc}.  By \Eq{eq:constraintksphere}, 
this can be done by taking $T_{kl} = 0$ along $N_{-k}(\sigma') \cap 
N_{+k}(Z)$.  Strictly speaking, we have thus far minimized $\Delta A_1$ 
by optimizing each free term of definite sign in the Raychaudhuri and 
cross-focusing equations in \Eq{eq:constraintk}, which is consistent 
with taking $Z$ to be a surface of constant affine parameter.  The 
remaining term in the cross-focusing equation, ${\cal D} \cdot \omega$, 
has indefinite sign and one could a~priori imagine using this term to 
bring $Z$ closer to $\sigma'$ along some generators of $N_{-k}(\sigma')$. 
However, taking $\omega$ to be nonzero along $N_{-k}(\sigma')$ 
requires turning on $({\cal D} \cdot \varsigma_k)_i$ or $T_{ik}$ 
by the Damour-Navier-Stokes equation, which in turn implies positive 
$\varsigma_k^2$ or $T_{kk}$, which take $\delta A(\nu)$ away from 
its optimal profile \eqref{eq:simple-resc}.  Moreover, since ${\cal D} 
\cdot \omega$ integrates to zero over any slice of $N_{-k}(\sigma')$, 
taking this term to be nonzero shifts some areas of $Z$ closer to 
$\sigma'$ and some farther away, in a manner that averages to zero 
for small $\omega$.  Since $\delta A(\nu)$ is convex in $\nu$, 
integrating $\delta A(\nu)$ over the angular directions for a 
distribution of $\nu$ values averaging to $\bar \nu$ always gives 
a smaller quantity than integrating $\delta A(\bar \nu)$ for constant 
$\bar \nu$.  Hence, a nonzero ${\cal D} \cdot \omega$ term only 
increases $\Delta A_1$, so our procedure thus far has indeed achieved 
the minimum value of $\Delta A_1$ consistent with our energy conditions 
and spherical symmetry of $\sigma'$.

We next consider $\Delta A_2$.  The constraint equations in 
\Eq{eq:constraintl} imply that we can achieve the optimal configuration 
of $\Delta A_2 = 0$ by taking $Y = Z$, so that $\theta_l$ vanishes 
at $Y$, and setting $\varsigma_l = T_{ll} = 0$ along $N_{+l}(Y)$ 
until we reach a surface with $\theta_k = 0$.  That is, we hold constant 
affine parameter slices of $N_{+l}(Y)$ to be stationary, so that 
each slice has the same area, while keeping $\omega_i = T_{kl} = 0$. 
This part of our setup is the time-reversed and $k \leftrightarrow l$ 
analogue of the construction in \Ref{Engelhardt:2017aux}.

We have now minimized $\Delta A_1$ and $\Delta A_2$ simultaneously, 
producing the HRT surface of maximal area consistent with the outer 
wedge for spherically-symmetric $\sigma'$.  The generality of the 
argument implies that this construction is indeed optimal.  Since the 
construction is precisely what we followed in deriving \Eq{eq:lower} 
in \Sec{subsec:construction}, the inequality there is in fact an 
equality:
\be
  \So{\sigma'} = \frac{A[\sigma']}{4G\hbar \xi_0^{D-2}}.
\label{eq:SOSequality}
\ee
In particular, this implies that any successful algorithm for 
maximizing the area of the HRT surface, not necessarily that 
of \Sec{subsec:construction}, would be guaranteed to reproduce 
\Eq{eq:SOSequality}.%
\footnote{For example, had we instead followed $N_{+l}(\sigma')$ to a 
 surface $X'$ on which $\theta_k = 0$ and then followed $N_{-k}(X')$ to 
 an HRT surface, the optimal construction would have yielded a surface 
 of the same area as given by \Eqs{eq:areaX}{eq:SOSequality}; this 
 follows from the manifest symmetry of \Eqs{eq:areaX}{eq:SOSequality} 
 under swapping $k \leftrightarrow l$: $\xi_0$ is a zero of the 
 polynomial given in \Eq{eq:ffinal}, with coefficients given in 
 \Eq{eq:rholambdadef} that are invariant under $k\leftrightarrow l$.}

We emphasize that $\xi_0$ in \Eq{eq:SOSequality} can be computed entirely 
from geometrical data on $\sigma'$.  We therefore have a new entry in 
the holographic dictionary: the spherical outer entropy of $\sigma'$ 
is a holographic quantity defined by the geometry of this leaf of 
the generalized holographic screen.  The outer entropy expression in 
\Eq{eq:SOSequality} is one of the main results of this work, giving 
an entropic interpretation to the generalized holographic screen.  This 
is especially interesting in the case in which $\sigma'$ corresponds 
to the event horizon: \Eq{eq:SOSequality} provides the first valid 
interpretation of the event horizon in terms of an entropic, holographic 
quantity computable from the horizon geometry.

\subsection{Cases of Interest}
\label{subsec:cases}

Though it is not possible to obtain an analytic expression for $\xi_0$ 
from \Eq{eq:ffinal} in complete generality, we can compute it in several 
cases of interest.  The first is the case of negligible $\lambda$, which 
corresponds to three possible situations: (i) an asymptotically-flat 
spacetime with $\Lambda = 0$, (ii) a black hole in which ${\cal R} \gg 
|\Lambda|$, i.e., a black hole much smaller than the (A)dS scale, and 
(iii) folding $\Lambda$ into $T_{ab}$ and, instead of the $\Lambda$DEC 
requirement, simply requiring the DEC on this entire $T_{ab}$.  Another 
case of interest is that of negligible $\rho$, corresponding to a black 
hole much larger than the length scale of the cosmological constant. 
Other particular situations to consider are three-dimensional spacetimes 
and surfaces in pure (A)dS or Minkowski space.  We will compute 
$\So{\sigma'}$ for each of these cases in turn.

\subsubsection[Small $\Lambda$]{Small {\boldmath $\Lambda$}}
\label{subsubsec:small-L}

Let us first consider the case in which $\Lambda$ is negligible in the 
polynomial in \Eq{eq:ffinal}, i.e., cases~(i), (ii), or (iii) above. 
We can then drop $\lambda$, so the zero in $q$ occurs at
\be
  \xi_0 = (1-\rho^{-1})^{-\frac{1}{D-3}}.
\label{eq:xi0flat}
\ee
Note that $\rho \rightarrow \infty$ corresponds to the apparent horizon, 
where $\theta_k[\sigma'] \rightarrow 0$.  Since ${\cal R}[\sigma'] > 0$, 
$\theta_k[\sigma'] > 0$, and $\theta_l[\sigma'] < 0$, we have 
$\rho > 0$.  Moreover, the condition \eqref{eq:rhopluslambda} 
for the zero requires $\rho > 1$ (which is automatically satisfied 
for a spherically-symmetric normal surface), so $\xi_0 > 1$.  Therefore, 
for generalized holographic screens with a geometry on $\sigma'$ 
satisfying $\rho > 1$, the spherical outer entropy is
\be
  \So{\sigma'} = \frac{A[\sigma']}{4G\hbar} 
    \left( 1 - \frac{1}{\rho} \right)^{\frac{D-2}{D-3}}.
\label{eq:SOSflat}
\ee
This provides us with an explicit entropic formula for the geometry of a 
generalized holographic screen, including the event horizon, for any outer 
wedge associated with a spherically-symmetric normal surface on which 
the cosmological constant is negligible.  It is then straightforward 
to compute $\rho$ for various spacetimes of interest and substitute 
into \Eq{eq:SOSflat} to yield the outer entropy.

\subsubsection[Large $\Lambda$]{Large {\boldmath $\Lambda$}}
\label{subsubsec:large-L}

Let us now consider the opposite limit, in which the cosmological 
constant dominates over the intrinsic curvature of the generalized 
holographic screen.  Since our construction in \Sec{subsec:construction} 
required $\rho + \lambda > 1$, in the limit in which $\Lambda$ dominates 
we must consider a negative cosmological constant $\Lambda < 0$ in 
order to have $\lambda>0$ (by \Eq{eq:rholambdadef}, recalling that 
$\theta_k[\sigma']>0$ and $\theta_l[\sigma']<0$), so we are in an 
asymptotically-AdS spacetime.  We consider a black hole much larger 
than the AdS length.  In this case, we can drop $\rho$ from the 
polynomial in \Eq{eq:ffinal} and solve for $\xi_0$:
\be
  \xi_0 = (1-\lambda^{-1})^{-\frac{1}{D-1}}.
\label{eq:xi0AdS}
\ee
Note that $\lambda \rightarrow \infty$ corresponds to the apparent 
horizon, $\theta_k[\sigma']\rightarrow 0$, for fixed $\Lambda$.  The 
condition~\eqref{eq:rhopluslambda} for the zero requires $\lambda > 1$, 
so $\xi_0 > 1$.  We thus have the outer entropy given by the geometry 
on $\sigma'$ in the large black hole limit:
\be
  \So{\sigma'} = \frac{A[\sigma']}{4G\hbar} 
    \left( 1 - \frac{1}{\lambda} \right)^{\frac{D-2}{D-1}}.
\label{eq:SOSAdS}
\ee
This is an entropic dual of the geometry of the generalized holographic 
screen, including the event horizon, for a black hole large compared to 
the AdS scale.

\subsubsection[$D=3$]{\boldmath $D=3$}
\label{subsubsec:3D}

If $D=3$, the analysis above needs to be modified.  In particular, in 
three spacetime dimensions, the polynomial in \Eq{eq:ffinal} becomes
\be
  q(\nu) = (1-\lambda)\xi(\nu)^2 + \lambda, 
\ee
so the terms involving $\rho$ cancel.  Note that, despite the factor 
of $D-3$ in the denominator of $\rho$ in \Eq{eq:rholambdadef}, there is 
also a factor of $D-3$ in the numerator arising from the intrinsic Ricci 
curvature given in \Eq{eq:Rdefsph}, so the cancellation of $\rho$ is 
well defined.  We therefore have
\be
  \xi_0 = \left(1-\lambda^{-1}\right)^{-\frac{1}{2}},
\ee
so that the solution behaves like the $\Lambda$-dominated case of 
\Sec{subsubsec:large-L}.  This implies that a surface with $\theta_l = 0$ 
can only be reached in $D=3$ for $\Lambda < 0$.

We can understand what is happening here from the cross-focusing equation 
for $\nabla_k \theta_l$ in \Eq{eq:constraintksphere}.  Even without 
assuming spherical symmetry, $\cal R$ vanishes in $D = 3$, since $\sigma'$ 
is simply a curve, which does not have intrinsic curvature.  Hence, 
the only term in \Eq{eq:constraintksphere} that can be negative---and 
thus allow $\theta_l$ to reach zero somewhere on $N_{-k}(\sigma')$---is 
$\Lambda$.  This requirement of negative cosmological constant 
accords with the fact that in $D = 3$ there are no black holes in 
asymptotically-flat or asymptotically-dS spacetimes, but there do exist 
BTZ black holes in asymptotically-AdS spacetimes~\cite{Banados:1992wn}.

\subsubsection{Vanishing entropy for (A)dS}
\label{subsubsec:AdS}

Suppose that $\sigma'$ is in a region of pure AdS for a black hole formed 
from collapse; for example, $\sigma'$ can be in the innermost region of 
AdS-Vaidya spacetime.  In \Ref{Engelhardt:2017wgc}, this spacetime was 
given as a counterexample to show that the area of the causal surface 
in this region cannot have a straightforward holographic interpretation 
as a von~Neumann entropy.  This conclusion follows from rigidity of the 
bulk vacuum, which implies that any spacetime one can construct with 
$O_W(\sigma')$ fixed would have no HRT surface for $\sigma'$ located 
in a pure AdS region.

We can see how the expression of our entropy in \Eq{eq:SOSequality} 
remains consistent in this setup.  For pure AdS spacetime,%
\footnote{For the straightforward extension to dS spacetime, one can 
 simply take $L^2$ to be negative.}
the metric is given by
\be
  ds^2 = -\biggl( 1+\frac{r^2}{L^2} \biggr) dt^2 
    + \frac{1}{1+\frac{r^2}{L^2}}dr^2 + r^2 d\Omega_{D-2}^2,
\label{eq:metric}
\ee
where
\be
  \Lambda = -\frac{(D-1)(D-2)}{2L^2}.
\label{eq:Lambdadef}
\ee
For the radial null vectors $k$ and $l$, with $k\cdot l = -1$, we can 
choose the relative normalization to be equal:
\be
  k^a, l^a = \frac{1}{\sqrt{2}} 
    \left( \frac{1}{\sqrt{1+\frac{r^2}{L^2}}},\, 
      \pm \sqrt{1+\frac{r^2}{L^2}},\, \vec{0} \right).
\label{eq:klCC}
\ee
With this choice, $dr/d\nu = \sqrt{(1+(r^2/L^2))/2}$ and
\be
  \theta_k = -\theta_l = \frac{D-2}{\sqrt{2}r}\sqrt{1+\frac{r^2}{L^2}}.
\label{eq:thetasAdS}
\ee

From the definitions in \Eq{eq:rholambdadef}, along with 
Eqs.~\eqref{eq:Rdefsph}, \eqref{eq:Lambdadef}, and \eqref{eq:thetasAdS}, 
we find that for a spherically-symmetric leaf $\sigma'$ in a pure 
(A)dS region,
\be
\begin{aligned}
  \rho    &= \left(1+\frac{r_0^2}{L^2}\right)^{-1},
\\
  \lambda &= \frac{r_0^2}{L^2} \left(1+\frac{r_0^2}{L^2}\right)^{-1},
\end{aligned} 
\ee
where $r_0$ represents the location of $\sigma'$.  We thus find that 
for spacetimes locally AdS, dS, or Minkowski around $\sigma'$, spherical 
light sheets obey \Eq{eq:ffinal} with
\be
  \rho + \lambda = 1.
\label{eq:special}
\ee
In these special cases, $q(\nu)$ does not have a zero for $\xi(\nu) > 1$, 
since the requirement in \Eq{eq:rhopluslambda} is violated.  In 
particular, $\theta_l \rightarrow -\infty$ when $\theta_k \rightarrow 
+\infty$ as $\nu \rightarrow \nu_{\rm c}$, which corresponds to the 
light sheets converging to a point at $r = 0$.  This implies that 
there is no HRT surface, so $\So{\sigma'} = 0$.  Formally, setting 
$\rho + \lambda = 1$ in \Eq{eq:ffinal}, $q(\xi)$ becomes $\rho\xi^2 + 1 
- \rho$, which has no zero in $(1,\infty)$ for positive $\rho$; in this 
case, as $\xi \rightarrow \infty$ (as $\theta_k \rightarrow \infty)$, 
\Eq{eq:qdef} implies that $\theta_l \rightarrow -\infty$.  If we instead 
take the limit as $\rho + \lambda \rightarrow 1$, the zero satisfies 
$\xi_0 \rightarrow \infty$, so \Eq{eq:SOSequality} implies that 
$\So{\sigma'} \rightarrow 0$.  Thus, the outer entropy we derived in 
\Eq{eq:SOSequality} does not suffer from the problem that the causal 
holographic information (which was given simply by the area, i.e., 
\Eq{eq:SOSequality} without the $\xi_0^{D-2}$ factor) had encountered.

\subsection{The Second Law}
\label{subsec:2nd-law}

Let us now compute how $\So{\sigma'}$ changes as we evaluate it for 
different leaves $\sigma'(\tau)$ along the generalized holographic screen 
$H'$.  By definition, the outer wedges for consecutive leaves along 
$H'$ are nested, $O_W(\sigma'(\tau_1)) \supset O_W(\sigma'(\tau_2))$ 
for $\tau_1 < \tau_2$.  This implies that the spacetime region held 
fixed when we scan possible spacetimes in finding the HRT surface of 
maximal area becomes progressively smaller.  Since a maximum evaluated 
on consecutively larger domains can only grow, it follows that we 
should have $\nabla_\tau \So{\sigma'(\tau)} \geq 0$.  We will now see 
explicitly how this comes about for the spherical outer entropy given 
by \Eq{eq:SOSequality}, which will serve as a nontrivial check on our 
result.  Note that the area law computed for $H'$ in \Sec{subsec:area} 
does not a~priori guarantee a second law for \Eq{eq:SOSequality}, 
since $\So{\sigma'(\tau)}$ is {\it not} simply the area of $\sigma'$; 
instead, we will find that the increase in the area of $\sigma'(\tau)$, 
along with the behavior of $\xi_0(\tau)$, will combine to give a second 
law for $\So{\sigma'(\tau)}$.

Even though the root $\xi_0$ of the polynomial in \Eq{eq:ffinal} cannot 
be expressed in closed form for general $D$, $\rho$, and $\lambda$, 
we can still prove the second law for $\So{\sigma'(\tau)}$.  Recalling 
that the tangent vector along $H'$ is $h'^a = \alpha l^a + \beta k^a$, 
we have
\be
\begin{aligned}
  \nabla_\tau \log \So{\sigma'(\tau)} 
  &= \alpha \nabla_l \log \So{\sigma'} + \beta \nabla_k \log \So{\sigma'}
\\
  &= \alpha [ \theta_l - (D-2) \nabla_l \log\xi_0 ] 
    + \beta [ \theta_k - (D-2) \nabla_k \log\xi_0 ],
\label{eq:derSOSgen}
\end{aligned}
\ee
where for the rest of this section, we will suppress the implicit argument 
of $\sigma'(\tau)$ in variables on the right-hand side.  Let us take the 
$\nabla_k$ derivative of
\be
  q(\nu_0) = (1-\rho-\lambda)\xi_0^{D-1} + \rho \xi_0^2 + \lambda = 0
\label{eq:xi0zero}
\ee
to get
\be
\begin{aligned}
  &\left(1-\rho-\lambda\right) 
    \left[(D-3)\rho\xi_0^2+(D-1)\lambda\right] \nabla_k \log\xi_0 
\\
  &\qquad = \left[\xi_0^{2}+\lambda(1-\xi_0^2)\right] \nabla_k \rho 
    + \left[1-\rho\left(1-\xi_0^2\right)\right] \nabla_k \lambda,
\label{eq:del_ks}
\end{aligned}
\ee
where we have used the condition \eqref{eq:xi0zero} again to write 
$\xi_0^{D-1}$ in terms of $\xi_0^2$.  The analogous equation also holds 
for the $\nabla_l$ derivative.

From the definitions in \Eq{eq:rholambdadef}, using the constraint 
equations in \Eq{eq:constraintksphere}
along with \Eq{eq:Rconst}, we find
\be
\begin{aligned}
  \nabla_k \rho &= \rho \left( {\cal R}^{-1} \nabla_k {\cal R} 
    - \theta_k^{-1}\nabla_k\theta_k - \theta_l^{-1}\nabla_k\theta_l 
    \right)
\\
  &= \rho \left( \frac{D-3}{D-2} \theta_k + \frac{{\cal R}}{2\theta_l} 
    - \frac{\Lambda}{\theta_l} + \frac{8\pi G}{\theta_k} T_{kk} 
    - \frac{8\pi G}{\theta_l} T_{kl} \right)
\\
  &= \rho \left[ \frac{D-3}{D-2} \theta_k (1-\rho) 
    - \frac{D-1}{D-2}\theta_k\lambda + 8\pi G\left(\frac{T_{kk}}{\theta_k} 
    - \frac{T_{kl}}{\theta_l}\right) \right]
\end{aligned}
\ee
and
\be 
\begin{aligned}
  \nabla_k \lambda &= -\lambda \left( \theta_k^{-1} \nabla_k \theta_k 
    + \theta_l^{-1} \nabla_k \theta_l \right)
\\
  &= \lambda \left( \frac{D-1}{D-2}\theta_k + \frac{{\cal R}}{2\theta_l} 
    - \frac{\Lambda}{\theta_l} + \frac{8\pi G}{\theta_k} T_{kk} 
    - \frac{8\pi G}{\theta_l} T_{kl} \right)
\\
  &= \lambda \left[ \frac{D-1}{D-2} \theta_k (1-\lambda) 
    - \frac{D-3}{D-2}\theta_k\rho + 8\pi G \left(\frac{T_{kk}}{\theta_k} 
    - \frac{T_{kl}}{\theta_l}\right) \right].
\end{aligned}
\ee
Hence, from \Eq{eq:del_ks} we obtain, again using the definition of the 
zero in \Eq{eq:xi0zero} and after some rearrangement,
\be
  \nabla_{k}\log\xi_0 = \frac{8\pi G}{\theta_k \theta_l}\, 
    \frac{\xi_0^{D-1}}{(D-3)\rho\xi_0^2+(D-1)\lambda} 
    \left(T_{kl}\theta_{k}-T_{kk}\theta_{l}\right) 
    + \frac{\theta_{k}}{D-2}.
\label{eq:derxi0k}
\ee
Using this relation and the analogous one for $\nabla_l$, 
\Eq{eq:derSOSgen} becomes
\be
\begin{aligned}
  \nabla_\tau \So{\sigma'(\tau)} 
  &= -\frac{8\pi G(D-2)\xi_0^{D-1}\So{\sigma'(\tau)}}
    {\theta_k \theta_l [(D-3)\rho\xi_0^{2}+(D-1)\lambda]} 
    \left[ (\alpha\theta_l + \beta\theta_k)T_{kl} - \alpha T_{ll} \theta_k 
    - \beta T_{kk} \theta_l \right]
\\
  &= -\frac{2\pi (D-2)\xi_0 A[\sigma']}
    {\hbar \theta_k \theta_l [(D-3)\rho\xi_0^{2}+(D-1)\lambda]} 
    \left[ (\alpha\theta_l + \beta\theta_k)T_{kl} - \alpha T_{ll} \theta_k 
    - \beta T_{kk} \theta_l \right].
\end{aligned}
\label{eq:derSOSgenfull}
\ee

Let us consider the signs of the factors appearing in 
\Eq{eq:derSOSgenfull} in turn.  The term in brackets in the 
denominator, $(D-3)\rho\xi_0^2 + (D-1)\lambda$, is guaranteed to be 
positive by \Eq{eq:rholambdacond2}.  Moreover, by \Eq{eq:alphabeta}, 
$\alpha\theta_l + \beta\theta_k > 0$.  In particular, we have 
$\alpha < 0$ and $\theta_l < 0$ on $\sigma'$ from the definition 
of a generalized holographic screen given in \Sec{subsec:def}, while 
$\beta > 0$ and $\theta_k > 0$ since we are considering the outermost 
spacelike portion of $H'$.  Together with $\xi_0 > 1$, we thus conclude 
that the entire prefactor in front of the last set of square brackets 
in \Eq{eq:derSOSgenfull} is positive.  Now, the NEC requires that 
$T_{kk}$ and $T_{ll}$ are both nonnegative, while the $\Lambda$DEC 
implies that $T_{kl} \geq 0$.  Thus, all the terms in the last set 
of square brackets in \Eq{eq:derSOSgenfull} are nonnegative.  This 
proves that the outer entropy given in \Eq{eq:SOSequality} obeys 
the second law of thermodynamics,
\be
  \nabla_\tau \So{\sigma'(\tau)} \geq 0,
\label{eq:secondlaw}
\ee
along the generalized holographic screen.  Interestingly, 
\Eq{eq:derSOSgenfull} is reminiscent of a Clausius relation, 
with $dS \propto dQ$ for some flow of energy-momentum.

\section{Conclusions}
\label{sec:concl}

In this work, we identified a large new class of codimension-one 
surfaces, the generalized holographic screens, that extend the concept 
of holographic screens~\cite{Bousso:2015mqa} to surfaces that are not 
marginally trapped.  The family of generalized holographic screens 
connect the concept of holographic screens with event horizons, as 
both are members of this larger class of geometric objects.  We showed 
in \Sec{sec:GHS} that all generalized holographic screens satisfy an 
area theorem~\eqref{eq:arealawfinal}, thus relating the previously 
known area theorems of \Ref{Hawking:1971tu} and \Ref{Bousso:2015qqa} 
(as well as the related area laws of Refs.~\cite{Nomura:2018kji,%
Krolak1982,Ashtekar:2004cn,Booth:2005qc}).

Further, we showed in \Secs{sec:outer}{sec:sphere} that 
generalized holographic screens have an entropic interpretation. 
In \Eq{eq:SOSequality}, we calculated the outer entropy---the 
largest von~Neumann entropy, computed via the HRT formula, for fixed 
outer wedge---for leaves of the generalized holographic screen for 
spherically-symmetric spacetimes and subsequently showed that this 
entropy obeys the second law of thermodynamics.

The interpretation of the event horizon geometry through some 
relation to the von~Neumann entropy---via a well defined holographic 
prescription---has hitherto been unknown in AdS/CFT.  In this paper, 
we have found such a connection, expressing a particular geometric 
quantity defined on the event horizon---notably, not simply the 
area---in terms of the outer entropy.  This outer entropy gives 
the maximum area of the HRT surface for the collection of geometries 
with fixed causal wedge; equivalently, this expresses the maximal 
entanglement entropy between the two sides of the black hole for 
a pure boundary state.

We note that the specific details of the construction of the 
generalized holographic screen in \Sec{sec:GHS} are in fact not 
necessary to obtain the area law result in \Eq{eq:arealawfinal} 
or the second law result in \Eq{eq:secondlaw}.%
\footnote{We thank Raphael Bousso for discussion on this point.}
Instead, it is sufficient to require that the outer wedges of 
infinitesimally separated leaves $\sigma'(\tau)$ be nested in the 
outer spacelike direction ($\alpha < 0$ and $\beta > 0$) and that 
$\sigma'(\tau)$ is a normal surface ($\theta_k > 0$ and $\theta_l < 0$). 
This is possible, e.g., even if $\sigma'(\tau)$ is not entirely within 
$N_{-k}(\sigma)$ for some single $\sigma \subset H$ as required for 
a general holographic screen in \Sec{sec:GHS}.  In fact, a weaker 
set of conditions guaranteeing $\alpha\theta_l + \beta\theta_k > 0$ 
is sufficient to obtain the area law of \Eq{eq:arealawfinal}, while 
the second law of \Eq{eq:secondlaw} requires the related condition 
of positivity of \Eq{eq:derSOSgenfull}.  A related example is 
the monotonicity theorem for renormalized leaf areas given in 
Ref.~\cite{Nomura:2018kji}.

\begin{figure}
\centering
  \includegraphics[width = 16cm]{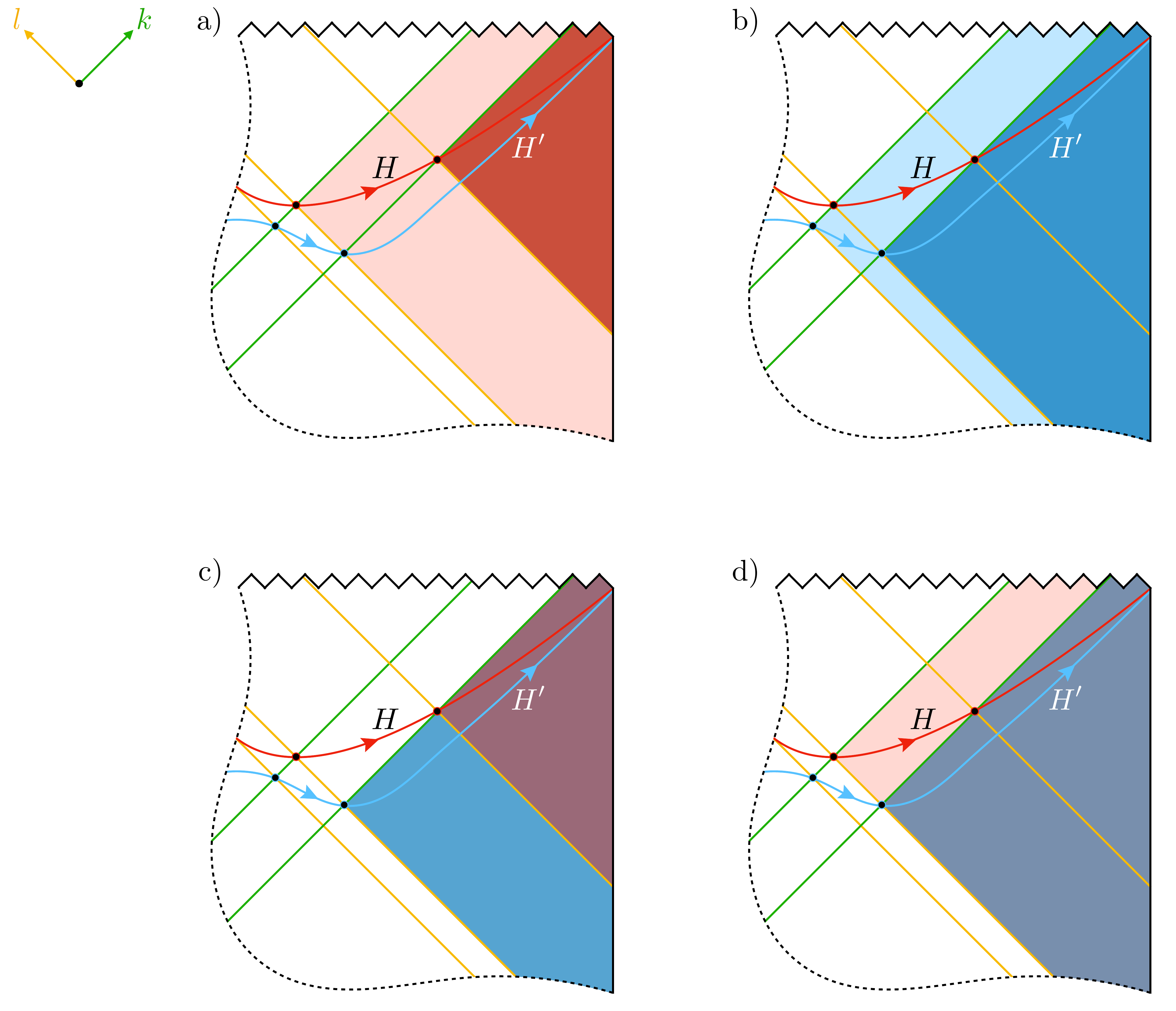}
\caption{Generic Penrose diagrams illustrating the relationship 
 between the outer wedges of the holographic screen $H$ (red line) 
 and generalized holographic screen $H'$ (blue line), in the spacelike 
 case.  In the direction of increasing $\tau$ (arrows), outer wedges 
 of leaves of $H$ are nested, as shown in panel a).  Similarly, 
 wedges of $H'$ are nested as $\tau$ increases, as shown in panel b). 
 This nesting mandates an increase in outer entropy on $H$ and 
 $H'$.  For spacelike screens, each leaf $\sigma' \subset H'$ is in 
 $N_{-k}(\sigma)$ for some leaf $\sigma \subset H$, leading to the 
 nesting $O_W(\sigma') \supset O_W(\sigma)$ illustrated in panel c). 
 In the case of a generalized holographic screen constructed via 
 intersections as in \Sec{subsec:intersection}, the opposite nesting 
 also occurs, as shown in panel d).  Outer wedges attached to leaves on 
 $H$ ($H'$) are shown in translucent red (respectively, blue), with 
 darker shades indicating increasing $\tau$.}
\label{fig:wedges}
\end{figure}

The generalized holographic screen $H'$ and holographic screen $H$ 
are related to each other by a network of coarse- and fine-graining 
relationships.  As illustrated in panels a) and b) of \Fig{fig:wedges}, 
the second law on $H$ associated with increase of the outer entropy 
can be understood from the nesting of outer wedges of leaves $\sigma 
\subset H$, i.e., coarse-graining of the data held fixed in the 
direction of increasing $\tau$, and similarly for $H'$.  Meanwhile, 
each leaf $\sigma' \subset H'$ is by definition in $N_k(\sigma)$ for 
some leaf $\sigma \subset H$.  For spacelike $H'$, $\sigma' \subset 
N_{-k}(\sigma)$ and we can therefore view the process of going from 
$H$ to $H'$ as a fine-graining (i.e., more data is being held fixed), 
since $O_W(\sigma') \supset O_W(\sigma)$, as shown in panel c) 
of \Fig{fig:wedges}, illustrating the upper bound $\So{\sigma'} \leq 
A[\sigma]/4G\hbar$.  Finally, in the case of a spacelike generalized 
holographic screen formed via the intersection construction of 
\Sec{subsec:intersection}, for each leaf $\sigma' \subset H'$ 
there is also a leaf in $H$ for which $\sigma'$ is on the $-l$ 
light sheet and for which the outer wedge contains $O_W(\sigma')$, 
as illustrated in panel d) of \Fig{fig:wedges}; in this direction, 
going from $H$ to $H'$ can be viewed as a coarse-graining.  In this 
case, the area of the corresponding leaf on $H$ provides a lower 
bound on $\So{\sigma'}$.

This work leaves numerous avenues for future research.  Investigation 
of the explicit boundary formulation of the outer entropy for 
non-marginally-trapped surfaces, in terms of boundary operators 
(cf.\ \Ref{Engelhardt:2017aux}) and the boundary density matrix, 
could prove fruitful.  Moreover, it would be very interesting to 
explore the meaning and utility of the outer entropy of generalized 
holographic screens in more general spacetimes as a compelling 
geometric quantity in the context of classical general relativity.

\vspace{5mm}
 
\begin{center} 
{\bf Acknowledgments}
\end{center}

\noindent 
We thank Ning Bao, Raphael Bousso, Sean Carroll, Netta Engelhardt, 
Illan Halpern, and Pratik Rath for useful discussions and comments. 
The work of Y.N.\ was supported in part by the National Science Foundation 
under grant PHY-1521446, by the Department of Energy, Office of Science, 
Office of High Energy Physics under contract No.\ DE-AC02-05CH11231, 
and by MEXT KAKENHI Grant Number 15H05895.  G.N.R. is supported by the 
Miller Institute for Basic Research in Science at the University of 
California, Berkeley.

\newpage

\bibliographystyle{utphys-modified}
\bibliography{GHS}

\end{document}